\documentstyle[12pt]{article}
\def\beq{\begin{equation}}
\def\eeq{\end{equation}}
\def\bea{\begin{eqnarray}}
\def\eea{\end{eqnarray}}
\def\bq{\begin{quote}}
\def\eq{\end{quote}}
\def\lesssim{\mathrel{\mathpalette\vereq<}}
\def\gtrsim{\mathrel{\mathpalette\vereq>}}
\makeatletter
\def\vereq#1#2{\lower3pt\vbox{\baselineskip1.5pt \lineskip1.5pt
\ialign{$\m@th#1\hfill##\hfil$\crcr#2\crcr\sim\crcr}}}
\makeatother\def\vereq#1#2{\lower3pt\vbox{\baselineskip1.5pt
\lineskip1.5pt
\ialign{$\m@th#1\hfill##\hfil$\crcr#2\crcr\sim\crcr}}}
\makeatother

\renewcommand{\thefootnote}{\fnsymbol{footnote}}
\begin{document}

\begin{titlepage}
\begin{center}
\today     \hfill    SLAC-PUB-7864\\
~{} \hfill SU-ITP-98/142\\
~{} \hfill IC/98/44\\
~{} \hfill hep-ph/9807344\\

\vskip .1in

{\large \bf
Phenomenology, Astrophysics and Cosmology\\
of Theories with Sub-Millimeter Dimensions\\
and TeV Scale Quantum Gravity}
\vskip 0.1in

Nima Arkani-Hamed$^a$, Savas Dimopoulos$^b$ and
Gia Dvali$^c$

\vskip .05in
{\em $^a$ SLAC, Stanford University,
Stanford, CA 94309, USA}
\vskip 0.1truecm
{\em $^b$ Physics Department, Stanford University,
Stanford, CA 94305, USA}
\vskip 0.1truecm
{\em $^c$ ICTP, Trieste, 34100, Italy}
\end{center}

\begin{abstract}
We recently proposed a solution to the hierarchy problem not relying
 on low-energy supersymmetry or technicolor. Instead,
the problem is nullified by bringing quantum gravity down to the
TeV scale. This is accomplished by the presence of $n \geq 2$
new dimensions of sub-millimeter size, with the SM fields
localised on a 3-brane in the higher dimensional space. In this
paper we systematically study the experimental viability of this
scenario. Constraints arise both from strong quantum gravitational
effects at the TeV scale, and more importantly from the production
of massless higher dimensional gravitons with TeV suppressed
couplings. Theories with $n>2$ are safe due mainly to the infrared
softness of higher dimensional gravity. For $n=2$, the six dimensional
Planck scale
must be pushed above $\sim 30$ TeV to avoid cooling SN1987A
and distortions of the diffuse photon background.
Nevertheless, the particular implementation of our framework within
type I string theory can evade all constraints, for any $n \geq
2$, with  string scale $m_s \sim 1$
TeV. We also explore novel phenomena resulting from the existence
of new states propagating in the higher dimensional space. The
Peccei-Quinn solution to the strong CP problem is revived with a
weak scale axion in the bulk. Gauge fields in the bulk
can mediate repulsive forces $\sim 10^6 - 10^8$ times stronger than gravity at
sub-mm distances, as well as help stabilize the proton.
Higher-dimensional gravitons produced on our
brane and captured on a different ``fat" brane can provide a
natural dark matter candidate.

\end{abstract}

\end{titlepage}
\def\simlt{\stackrel{<}{{}_\sim}}
\def\simgt{\stackrel{>}{{}_\sim}}
\newcommand{\cm}{Commun.\ Math.\ Phys.~}
\newcommand{\prl}{Phys.\ Rev.\ Lett.~}
\newcommand{\pr}{Phys.\ Rev.\ D~}
\newcommand{\pl}{Phys.\ Lett.\ B~}
\newcommand{\ibar}{\bar{\imath}}
\newcommand{\jbar}{\bar{\jmath}}
\newcommand{\np}{Nucl.\ Phys.\ B~}
\newcommand{\be}{\begin{equation}}
\newcommand{\en}{\end{equation}}
\newcommand{\ba}{\begin{eqnarray}}
\newcommand{\ea}{\end{eqnarray}}
\newcommand{\aG}{\alpha_G}

\def\lesssim{\mathrel{\mathpalette\vereq<}}
\def\gtrsim{\mathrel{\mathpalette\vereq>}}
\makeatletter
\def\vereq#1#2{\lower3pt\vbox{\baselineskip1.5pt \lineskip1.5pt
\ialign{$\m@th#1\hfill##\hfil$\crcr#2\crcr\sim\crcr}}}
\makeatother

\renewcommand{\thefootnote}{\fnsymbol{footnote}}

\def\beq{\begin{equation}}   \def\eeq{\end{equation}}

\newcommand{\gsim}{\lower.7ex\hbox{$\;\stackrel{\textstyle>}{\sim}\;$}
}
\newcommand{\lsim}{\lower.7ex\hbox{$\;\stackrel{\textstyle<}{\sim}\;$}
}

\newcommand{\ra}{\rightarrow}
\newcommand{\ve}[1]{\vec{\bf #1}}

\newcommand{\La}{\overline{\Lambda}}
\newcommand{\Lam}{\Lambda_{QCD}}
\newcommand{\re}[1]{Ref.~\cite{#1}}

\section{Introduction}
In a recent paper \cite{ADD}, we have proposed a framework for solving
the
hierarchy problem which does not rely on supersymmetry or
technicolor. Rather, the problem is solved by removing its
premise: the fundamental Planck scale, where gravity becomes comparable
in strength to the
other interactions, is taken to be near the weak scale. The
observed weakness of gravity at long distances is due to the
presence of $n$ new spatial dimensions large compared to the
electroweak scale.
This can be inferred from the relation between the Planck scales of the
$(4+n)$ dimensional
theory $M_{Pl(4+n)}$ and the long-distance 4-dimensional theory
$M_{Pl(4)}$, which can simply be determined
by Gauss' law (see the next section for a more detailed
explanation)
\beq
M_{Pl(4)}^2 \sim r_n^n M_{Pl(4+n)}^{n+2}
\eeq
where $r_n$ is the size of the extra dimensions.
Putting $M_{Pl(4+n)} \sim 1$ TeV then yields
\beq
r_n \sim 10^{30/n - 17} \mbox{cm}
\eeq
For $n=1$, $r_1 \sim 10^{13}$ cm, so this case is
obviously excluded since it would modify Newtonian gravitation at
solar-system distances. Already for $n=2$, however, $r_2 \sim 1$
mm, which is precisely the distance where our present experimental
measurement of gravitational strength forces stops. As $n$
increases, $r_n$ approaches $($TeV$)^{-1}$ distances, albeit
slowly: the case $n=6$ gives $r_6 \sim (10$MeV$)^{-1}$. Clearly,
while the gravitational force has not been directly measured beneath
a millimeter, the success of the SM up to $\sim 100$ GeV implies
that the SM fields can not feel these extra large dimensions; that
is, they must be stuck on a wall, or ``3-brane", in the higher
dimensional space. Summarizing, in our framework the universe is $(4+n)$

dimensional with Planck scale near the weak scale, with $n \geq 2$
new sub-mm sized dimensions where gravity  perhaps other fields can
freely propagate,
but where the SM particles are  localised on a 3-brane
in the higher-dimensional space.

An important question is the mechanism by which the SM fields are
localised to the brane.
In \cite{ADD}, we proposed a field-theoretic implementation
of our framework based on earlier ideas for localizing the
requisite spin 0,1/2\cite{RS} and 1 \cite{DS} particles.
In \cite{AADD} we showed that our framework can naturally
be embedded in type I string theory. This has the obvious
advantage of being formulated within a consistent theory of
gravity, with the additional benefit
that the localization of gauge theories on a 3-brane is automatic 
\cite{Dbrane}. Further interesting 
progress towards realistic string model-building was made in \cite{Tye}.

The most pressing issue, however, is to insure that this framework
is not experimentally excluded. This is a concern for two main
reasons. First, quantum gravity has been brought down from
$10^{19}$ GeV to $\sim$ TeV. Second, the structure of
space-time has been drastically modified at sub-mm distances. The
main objective of this paper is to examine the
phenomenological, astrophysical and cosmological constraints on
our framework. Subsequently, we discuss a number of new phenomena
which emerge in theories with large extra dimensions.

The rest of the paper is organized as follows. In section 2, we
derive the exact relationship between the
Planck scales of the $(4+n)$ and 4-d theories in three ways in
order to gain some intuition for the physics of higher dimensional
theories. Of course, roughly speaking, if $M_{Pl(4+n)} \sim 1$
TeV, we expect new physics responsible for making a sensible
quantum theory of gravity at the TeV scale. 
There is a practical difference between the new physics occurring at $\sim 1$ TeV
versus $\sim 10$ TeV, as far as accessibility to future colliders is concerned.
In section 3, we therefore give a more careful account of the relationship between the scale of 
new physics 
and $M_{Pl(4+n)}$ in the particular case where gravity is embedded in type I 
string theory.
As a set-up for the discussion of phenomenological constraints,in
section 4
we identify and discuss the interactions of new light particles in the
effective theory
beneath the TeV scale: higher dimensional graviton, and possibly Nambu-Goldstone
bosons of broken translation invariance. In section 5 we
begin the discussion of phenomenological constraints in earnest, beginning 
with
laboratory experiments. 
The most stringent bounds are not due to
strong gravitational effects at $\sim$ TeV energies,
but rather due to the
possibility of producing massless particles, the higher dimensional
gravitons, 
whose couplings are only
1/TeV suppressed.  In section 5 we discuss potential problems
this can cause with rare decays, and in sections 6 and 7 we
consider astrophysical and cosmological constraints.
Remarkably, due primarily to the extreme infrared
softness of higher dimensional gravity, we find that for $n>2$ all
experimental limits are comfortably satisfied. The case $n=2$ is quite
tightly constrained, with a lower bound $\gsim 30$ TeV on the 6-d
Planck scale. Nevertheless, precisely for $n=2$, this Planck mass
can still be consistent with string excitations at the TeV scale,
and therefore may still provide a natural solution to the
hierarchy problem. Not only are cosmological constraints
satisfied, there are new cosmological possibilities in our scenario.
In particular, we discuss the possibility that 
gravitons produced on our brane and
captured on a different, ``fat" brane in the bulk, can form the dark matter of the Universe.
The following two sections illustrate further possibilities for
new physics in this framework. In  section 9, we  show that the
Peccei-Quinn axion can solve the Strong-CP problem and avoid the
usual astrophysical bounds if the axion field lives in the bulk.
In section 10, we note that a gauge field living in the bulk can
naturally have a miniscule gauge coupling $\sim 10^{-16}$ to wall states
and pick
up a mass $\sim 1$mm$^{-1}$ through spontaneous breaking on the wall. If these gauge fields couple to $B$
or $B-L$, they can mediate repulsive forces $\sim 10^{6}-10^{8}$ 
times stronger than
gravity at sub-mm distances. This gives a spectacular experimental signature
may be observed in the near future. Finally, in section 11, we
turn to the important question of the determination of the radii of
the extra dimensions. While we do not offer any dynamical
proposal, we parametrize the potential for the radius modulus and
consider cosmological constraints coming from the requirement 
that the radius is not significantly altered since before the era of 
Big-Bang Nucleosynthesis (BBN). 
We draw our conclusions in section 12. Appendix 1 discusses the
somewhat subtle issue of the Higgs phenomenon for
spontaneously broken translational invariance, and appendix 2 presents a toy model illustrating some aspects 
of moduli stabilization.

\section{Relating Planck Scales}

\subsection{Gauss Law}
Here we will derive the exact relationship between the Newton constants
$G_{N(4+n)},G_{N(4)}$ of the full $(4+n)$ and compactified 4 dimensional
theories, which are defined by the force laws
\begin{eqnarray}
F_{(4+n)}(r) &=& G_{N(4+n)} \frac{m_1 m_2}{r^{n+2}} \nonumber \\
F_{(4)}(r) &=& G_{N(4)} \frac{m_1 m_2}{r^2}.
\end{eqnarray}
We will carry out this simple exercise in three different
ways. The easiest derivation is a trivial application of Gauss' Law.
Let us compactify the
$n$ new dimensions $y_\alpha$ by making the periodic
identification $y_\alpha \sim y_\alpha + L$.
Suppose now that a point mass $m$ is placed at the origin. One can
reproduce this situation in the uncompactified theory by placing
``mirror"
masses periodically in all the new dimensions.
Of course for a test mass at distances $r \ll L$ from $m$, the ``mirror"
masses
make a negligible small contribution to the force and we have the
$(4+n)$
dimensional force law. For $r \gg L$, on the other hand,
the discrete distance between mirror masses can not be discerned
and they look like
an infinite $n$ spatial dimensional ``line" with uniform mass density.
The problem is analogous to 
finding the gravitational field of an infinite line of mass with uniform
mass/unit length, where cylindrical symmetry and Gauss' law give the
answer.
Following exactly the same procedure, we consider a ``cylinder" $C$
centered around the $n$ dimensional line of mass, with side length $l$
and
end caps being three dimensional sphere's of radius $r$. We now apply
the $(4+n)$ dimensional Gauss' law which reads
\begin{equation}
\int_{\mbox{surface}\, C} F dS = S_{(3 + n)} G_{N(4+n)} \times
\mbox{Mass in} \, C
\end{equation}
where $S_{D} = 2 \pi^{D/2} / \Gamma(D/2)$
is the surface area of the unit sphere in $D$ spatial
dimensions (recall that the usual Gauss law has a $4 \pi$ factor on the
RHS).
In our case, the LHS is equal to $F(r) \times 4 \pi \times l^n$, while
the total
mass contained in $C$ is $m \times (l^n/L^n)$. Equating the two sides,
we find
the correct $1/r^2$ force law and can identify
\begin{equation}
G_{N(4)} = \frac{S_{(3+n)}}{4 \pi} \frac{G_{N(4+n)}}{V_n}
\label{GN}
\end{equation}
where $V_n = L^n$ is the volume of compactified dimensions.

We can also derive this result directly by compactifying the Lagrangian
from
$(4+n)$ to $4$ dimensions, from which we can also motivate a definition
for the
``reduced" Planck scale in both theories.
In the non-relativistic limit and in $(4+n)$ dimensions,
the action for the interaction of the (dimensionless)
gravitational potential $\phi = g^{00} - 1$,
with a mass density $\rho$, is given by
\begin{equation}
I_{(4+n)} = \int d^{4+n} x \frac{1}{2} \hat{M}_{(4+n)}^{n+2}
\phi \nabla_{(3+n)}^2 \phi + \rho_{(4+n)} \phi + \cdots
\end{equation}
where $\nabla^2_D$ is the $D$ spatial dimensional Laplacian,
and we define $\hat{M}_{(4+n)}$ as the reduced Planck scale in
$(4+n)$ dimensions.
Note that if we wish to work with canonically normalized $\phi$ field,
we rewrite $\phi = \hat{M}^{-1/2}_{(4+n)} \phi_{can}$, and the
Lagrangian
becomes
\begin{equation}
I_{(4+n)} = \int d^{4+n} x \frac{1}{2} \phi_{can} \nabla_{(3+n)}^2
\phi_{can} +
\frac{1}{\sqrt{\hat{M}_{(4+n)}^{n+2}}}
\rho_{(4+n)} \phi_{can} + \cdots
\end{equation}
showing that the interaction of the canonically normalized field are
suppressed
by $\sqrt{\hat{M}_{(4+n)}}$.
Upon integrating out $\phi$, we generate the potential
\begin{equation}
\int dt d^{(3+n)} x d^{(3+n)} y \frac{1}{\hat{M}_{(4+n)}^{n+2}}
\rho_{(4+n)}(x) \nabla^{-2}_{(3+n)}(x-y) \rho_{(4+n)}(y).
\end{equation}
Using
\begin{equation}
\nabla^{-2}_D (x-y) = \frac{1}{(D-2) S_D} \frac{1}{|x - y|^{D-2}}
\end{equation}
we have for the force between two test masses
\begin{equation}
F_{(4+n)}(r) = \frac{1}{\hat{M}_{(4+n)}^{n+2} S_{(3+n)}} \frac{m_1
m_2}{r^{n+2}}
\end{equation}
from which we find the relationship between the reduced Planck scale and
Newtons constant
\begin{equation}
\hat{M}_{(4+n)}^{n+2} = \frac{G_{N(4+n)}^{-1}}{S_{(3+n)}}.
\label{red}
\end{equation}
We can compactify from $(4+n)$ to 4 dimensions by restricting all the
fields
to be constant in the extra dimensions; integrating over the $n$
dimensions
then yields the 4 dimensional action
\begin{equation}
I_4 = \int d^4 x \frac{1}{2} (V_n \hat{M}_{(4+n)}^{n+2}) \phi \nabla_3^2
\phi
+ \rho_3 \phi + \cdots
\end{equation}
and therefore the reduced Planck scales of the two theories are related
according to
\begin{equation}
M_{(4)}^2 = \hat{M}_{(4+n)}^2 V_n,
\label{redp}
\end{equation}
which using eqn.(\ref{red}) reproduces the relation between the Newton
constants in eqn.(\ref{GN}).
An interesting string theoretic application of this result was made in \cite{HW},
where it was used to low the string scale to the GUT scale, choosing the radius of
11'th dimension in $M$-theory to be $\sim 10^{-27}$cm. Attempts 
to reduce the string scale much further were considered in \cite{Lower}, 
but their conclusions were basically negative.   

Finally, we can understand this result
purely from the 4-dimensional point of view as arising from the sum over
the
Kaluza-Klein excitations of the graviton. From the 4-d point of view,
a $(4+n)$ dimensional graviton with momentum $(q_1, \cdots, q_n)$ in the
extra
$n$ dimensions looks like a massive particle of mass $|q|$. Since the
momenta
in the extra dimensions are quantized in units of $2 \pi/L$,
this corresponds to an infinite tower of KK excitations for each of
the $n$
dimensions, with mass splittings $2 \pi/L$. While each of these KK modes
is very
weakly coupled ($\sim 1/M_{(4)}$), their large multiplicity can give a
large
enhancement to any effect they mediate. In our case, the potential
between two
test masses not only has the $1/r$ contribution from the usual
massless
graviton, but also has Yukawa potentials mediated by all the massive
modes as
well:
\begin{equation}
\frac{V(r)}{m_1 m_2} = G_{N(4)} \sum_{(k_1, \cdots,k_n)}
\frac{e^{-(2 \pi |k|/L) r}}{r}.
\end{equation}
Obviously, for $r \ll L$, only the ordinary massless graviton
contributes and we
have the usual potential. For $r \gg L$, however, roughly $(L/r)^n$ KK
modes
make unsuppressed contributions, and so the potential grows more rapidly
as
$L^n/r^{n+1}$. More exactly, for $r \gg L$,
\begin{eqnarray}
\frac{V(r)}{m_1 m_2} &\to&  G_{N(4)}{r} \times (\frac{L}{2 \pi R})^n
\times
\int d^n u e^{-|u|} \nonumber \\
&=& \frac{G_{N(4)} V_n}{r^{n+1}} \times \frac{S_n \Gamma(n)}{(2 \pi)^n}.
\end{eqnarray}
This yields the same relationship between $G_{N(4)}$ and $G_{N(4+n)}$
found in
eqn.(\ref{GN}) upon using the Legendre duplication formula
\begin{equation}
\Gamma(\frac{n}{2}) \Gamma(\frac{n}{2} + \frac{1}{2}) =
\frac{\sqrt{\pi}}{2^{n-1}} \Gamma(n).
\end{equation}
We will encounter this phenomenon repeatedly in this paper: the
interaction of
the higher dimensional gravitons can be understood in two ways.
Directly from the (4+n) dimensional point of view, the graviton
couplings
are suppressed $1/\sqrt{\hat{M}_{4+n}}$, which can be understood from
the $4$ dimensional point of view as arising from a sum over a
large multiplicity of KK excitations each of
which has couplings suppressed by $1/M_{(4)}$. Note that for higher $n$,
the couplings of $(4+n)$ dimensional gravitons are
suppressed by more powers of the $(4+n)$
dimensional Planck scale $\hat{M}_{(4+n)}$, and so their
interactions become increasingly soft in the infrared (the flip-side of
the worse UV problems!). As already mentioned, and as will be seen in
detail in
many examples, this IR softness is crucial to
the survival of the theory when the fundamental Planck scale is taken to
be
near the TeV scale.

We make one last comment on $2 \pi$ factors. If we express $V_n = L_n^n
=
(2 \pi r_n)^n$, the first KK excitation has a mass $ r_n^{-1}$,
and $r_n$ (not $L_n$) more correctly describes the physical size of the
extra dimensions. For instance, the potential between two test masses
a distance $L_n$ apart is only modified at $O(e^{-2 \pi}) \sim 10^{-2}$,
whereas the change is $O(1)$ for a distance $r_n$ apart. In terms of
$r_n$,
the relation eqn.(\ref{redp}) becomes
\begin{equation}
M_{(4)}^2 = M_{(4+n)}^{(n+2)} r_n^n, \, M_{(4+n)}^{n+2}
\equiv (2 \pi)^n \hat{M}_{(4+n)}^{n+2}.
\label{M4p}
\end{equation}
We will see below that the experimental bounds most directly constrain
$M_{(4+n)}$, and that it is $M_{(4+n)}$ which is
required to be close to the weak scale for solving the hierarchy
problem.
Putting in the numbers, we find for $r_n$
\beq
r_n = 2 \times 10^{31/n - 16} \mbox{mm} \times \left(\frac{1
\mbox{TeV}}{M_{(4+n)}}\right)^{1 + 2/n}
\label{rn}
\eeq
For $n=2$ and $M_{(6)}$= 1 TeV, $r_2 \sim 1$ mm, which is
precisely the distance at which
gravity is currently measured directly. For $n=6$, $r_6 \sim (10$
Mev$)^{-1}$,
and for very large $n$, $r_n$ approaches
$M_{(4+n)}^{-1}$.

\section{Relating the Planck scale to the String Scale}
In this subsection we wish to be more precise about the various scales
in our problem. Namely, we wish to quantify what exactly what we mean by

``gravity gets strong at the weak scale". Of course, we are really
interested in
relating the scale $m_{grav}$ at which the new physics responsible
for making a sensible quantum theory of gravity appears, to parameters
of the
low-energy theory such as e.g. $G_{N(4+n)}$ or $M_{(4+n)}$. There is a
practical reason for finding determining this relationship.
Both theoretically and experimentally, $m_{grav} \lsim 1$ TeV is most
desirable, on the other hand, the most stringent experimental
constraints
we will discuss directly constrain the interactions of the
$(4+n)$ dimensional gravitons and hence put a bound on
$M_{(4+n)}$. It is therefore important to determine how this bound
translates into a
constraint on $m_{grav}$.

Without a specific theory in mind, it is difficult to relate $m_{grav}$
to
$M_{(4+n)}$, other than the expectation that they are ``closeby".
To be more concrete, we suppose that the theory above $m_{grav}$ is a
string
theory, specifically the realization of our scenario within type $I'$
string theory outlined in \cite{AADD}, which we briefly review
here. The low-energy action of type-I string theory in
10-dimensions reads
\begin{equation}
S = \int d^{10} x \left( \frac{m_s^8}{(2 \pi)^7 \lambda^2} {\cal R} +
\frac{1}{4} \frac{m_s^6}{(2 \pi)^7 \lambda} F^2 + \cdots \right).
\end{equation}
where $\lambda \sim e^\phi$ is the string coupling, and $m_s$ is the
string
scale, which we can identify with $m_{grav}$.
Compactifying to 4 dimensions on a manifold of volume $V_6$, we can
identify the resulting coefficients of $R$ and $(1/4) F^2$ with
$M_{(4)}^2$
and $1/g^2_4$,  from which we can find
\begin{eqnarray}
M_{(4)}^2 = \frac{(2 \pi)^7}{V_6 m_s^4 g_4^2} \nonumber \\
\lambda= \frac{g_4^2 V_6 m_s^6}{(2 \pi)^7}.
\end{eqnarray}
Putting $m_s \sim 1$ TeV and $g_4 \sim 1$ fixes a very small
value for $\lambda$ and a compactification volume much {\it smaller}
than the
string scale.
A more appropriate description is obtained by $T-$ dualising,
where we compactify on a manifold of volume $V'_6$ with a new string
coupling $\lambda'$ given by
\begin{eqnarray}
V'_6 &=& \frac{(2 \pi)^{12}}{V_6 m_s^{12}}, \nonumber \\
\lambda' &=& \frac{(2 \pi)^6}{m_s^6 V_6} \lambda = \frac{g_4^2}{2 \pi}.
\end{eqnarray}
In this $T-$dual description, the KK excitations of the open strings
in the
type-I picture become winding modes of type-I' open strings stuck to a
D3
brane, while only the closed string (gravitational) sector propagates in
the bulk.
Thus our scenario for solving the hierarchy problem can naturally be embedded in this picture.
The 4-dimensional Planck scale
\begin{equation}
M_4^2 = \frac{2 \pi}{g_4^4} m_s^{8} \left(\frac{V_6}{(2 \pi)^6}\right)
\label{M4}
\end{equation}
can then be much larger than the string scale if $V_6$ is much bigger
than $m_s^{-6}$.
To make contact with our framework, we assume that of the six compact
dimensions,
$(6-n)$ have a size $L_{(6-n)} = (2 \pi r_{(6-n)})$ with the ``physical"
size
$r_{(6-n)} \sim m_s^{-1}$, while the remaining $n$ dimensions of size
$L_n = 2 \pi r_n$
are the ``large" ones we previously discussed. Then $V_6/(2 \pi)^6 =
r_{(6-n)}^{(6-n)} r_n^n$,
and combining eqns(\ref{M4},\ref{M4p}) we obtain
\begin{equation}
\frac{M_{(4+n)}}{m_s} = \left(\frac{2 \pi}{g_4^4}\right)^{\frac{1}{n+2}}
(r_{(6-n)} m_s)^{\frac{6-n}{n+2}}.
\label{msm}
\end{equation}
It is clear that for the higher values of $n$, the possible enhancements
of $M_{(4+n)}/m_s$ from the
first two factors are negligible and we should expect $M_{(4+n)} \sim
m_s \lsim 1$ TeV.
For the case $n=2$, however,  the first factor can range from 2 to 3
depending on which of the
SM gauge couplings are chosen to represent $g_4$, and we can choose
$r_4$ to be somewhat larger than the string scale perhaps as low as
$\sim (300$GeV$)^{-1}$.
These factors can be enough to push $M_{(4+n)}$ to somewhat higher
values
$\sim 10$ TeV while keeping $m_s \sim 1$ TeV.  As will see later, the
strongest constraints occur for the
lowest values of $n$ and in some cases will indeed push $M_{(4+n)}$ above
$\sim
10$ TeV. It is reassuring to know that even in
this case, new string physics may be seen at
$\sim 1$ TeV.

\section{Couplings of Bulk Gravitons and Nambu-Goldstones of Broken Space-Time Symmetries.}
In this section, we wish to describe the light degrees of freedom
which exist in the effective theory beneath the scale of quantum gravity $m_{grav}$ and the
tension $f$ of the wall. In our scenario it is most natural to
assume $f \sim m_{grav}$. This sort of effective theory is interesting
because some states (such as the SM fields) live on a wall in the
extra dimensions, while other fields (such as the
gravitons) can freely propagate in the higher dimensional space. Of
course, the presence of the wall breaks translational invariance
in the extra $n$ dimensions. 
Part of our discussion
depends on whether this is a spontaneous or explicit breaking of
the $(4+n)-$d Poincare symmetry.
Let the
position of a point $x$ on the wall, in the higher dimensions
$a=4,\cdots 3+n$, be given by $y^a (x)$. 
In the case where the breaking is
spontaneous, wall configurations which differ from each other by a
uniform translation $y^a(x) \to y^a(x) + c$ are degenerate in energy. The
$y^a(x)$ are then dynamical fields, Nambu-Goldstone bosons of
spontaneously broken translation invariance. The fields in the
effective theory consist of the $y^a(x)$, together with the SM
fields on the wall and gravity in the full higher dimensional bulk.
The interactions of this effective theory are constrained by the
requirement that the full $(4+n)-d$ Poincare invariance be realized
non-linearly on these fields. A very nice analysis of the structure of
this effective theory together with the leading terms in its energy
expansion has recently been given by Sundrum \cite{Raman}. We will
not repeat this analysis here, as many of the details are
unimportant for phenomenological constraints we consider. We will instead
study the form of the least suppressed interactions to the
$y^a$ and the bulk gravitons.

Before turning to this, we raise a  puzzling question not addressed in
\cite{Raman}. Since gravity can be
thought of as gauging translation invariance, and since
translation invariance is spontaneously broken, why are the $y^a(x)$
not ``eaten"  by the corresponding ``gauge field" $g^{\mu a}$,
which would
become massive?
We analyze this question in appendix 1. The conclusion is that the
$y^a(x)$ are indeed eaten and the corresponding 4-d ``gauge" field gets
a mass
$\sim f^2/M_{(4)} \sim$ (1 mm)$^{-1}$ for $f \sim 1$ TeV. Notice
that, if $M_{(4+n)}$ is held fixed and $r_n \to \infty$,
this mass goes to zero since $M_{(4)} \to \infty$, and so that the
analysis of \cite{Raman}, which was implicitly done in this limit,
is unaffected.
Furthermore,this mass is so small that almost processes we consider
will involve energies $\gg 1$(mm)$^{-1}$, and so by the
equivalence theorem, it is much more convenient to think in terms
of the original picture of massless gravitons and Nambu-Goldstone
fields. Nevertheless, as we will see later,
a $\sim$(mm)$^{-1}$ mass is generically be generated for any ``bulk"
gauge field when the gauge symmetry is broken on the wall,
and can lead to very interesting experimental consequences.

We now turn to the leading couplings, first to goldstone fields
ignoring gravity, then to gravity ignoring the goldstones.
To begin with, note that the $y^a$ have mass dimension $-1$,
and are therefore written in terms of the canonically normalized
goldstone fields
$\pi^a$ as $y^a = \pi^a/f^2$. This is in analogy to the usual case
of Goldstone boson of internal symmetries, where the analogue of $y^a$ is an angle
$\theta^a$
of the group transformation, related to the physical pion fields
as $\theta^a = \pi^a/f$. This immediately means that the
interactions with the $y^a$, for $f \sim 1$ TeV, are always weaker than
neutrino interactions which are suppressed only by $\sim 1/m_W^2$.
In fact, it is easy to see that for interactions with scalars or
vectors or a single Well fermion, the leading operators must
involve {\it two} $y's$ and are therefore even more suppressed
$\sim 1/f^4$. This follows from a completely straightforward
operator analysis, but can also be simply understood as follows.
The fluctuations in the wall given by $\partial_{\mu} y^{a}(x)$
induce a non-trivial metric on the wall, inherited from the metric of
the bulk
space. Ignoring gravity, the bulk metric is flat and the induced metric
on the wall is
\beq
g_{\mu \nu} = \eta_{\mu \nu} + \partial_{\mu} y^a \partial_{\nu}
y^a
\eeq
which is symmetric under $y^a \to -y^a$, so the interactions of $y$
which result from non-trivial $g$ involve pairs of $y's$.
Following \cite{Raman}, an operator involving a single $y$
interacting with vector-like Well fermions $(\psi,\psi^c)$ can also
be written
\beq
{\cal O}_{1 y}  = c \psi \partial^{\mu} \psi^c \partial_{\mu} y
\eeq
Of course, since this operator violates chirality, we expect that
the coefficient $c$ is suppressed by $\sim m_{\psi}/m_{grav}$,
up to a model-dependent coefficient.

Next consider the coupling of the SM fields to gravitation, but
without exciting the $y^a$. If the bulk metric is $G_{MN}$ where
$M,N=0,\cdots,3+n$, the
induced metric on the wall is trivially
\beq
g_{\mu \nu}(x) = G_{\mu \nu}(x,y^a=0).
\eeq
The bulk gravitons are the perturbations of $G_{MN}$ about $\eta_{MN}$,
\beq
G_{MN} = \eta_{MN} + \frac{H_{MN}}{\sqrt{M_{(4+n)}^{n+2}}}
\eeq
and the linear interactions with SM wall fields
are given by
\beq
\int d^4 x T^{\mu \nu} \frac{H_{\mu \nu}(x,
y^a=0)}{\sqrt{M_{(4+n)}^{n+2}}}
\eeq
where $T^{\mu \nu}$ is the 4-d energy momentum tensor for the SM
fields. Two things are immediately obvious from this coupling.
First, there is no coupling to the $H_{a \mu},H_{ab}$
gravitons. This is intuitively clear: without changing the shape
of the wall (i.e. exciting the $y^a$), the wall fields make zero
contribution to $T^{\mu a},T^{ab}$ and the the couplings to the
corresponding $H'$s vanish.
Second, the interaction clearly violates translation
invariance in the extra dimensions, and therefore the extra
dimensional momenta $p^a$ need not be conserved however in the
interactions between the wall and bulk states, while energy is
still
conserved because time translational invariance still holds.
More intuitively we can think of the wall as being
infinitely heavy, so that it can
recoil to absorb extra-dimensional momentum without absorbing
energy. This can also be seen explicitly by expanding $H_{\mu \nu}(x,
y^a=0)$
into Kaluza-Klein modes
\beq
H_{\mu \nu}(x, y^a=0) = \sum_{n^a} \frac{1}{\sqrt{r_n^n}} H^{n^a}_{\mu
\nu}
\eeq
which shows that the wall $T^{\mu \nu}$ couples to all KK modes
with equal strength $1/M_{(4)}$.
Of course, there are many other couplings involving combinations
of the $y^a$ and gravitons, but they are all suppressed by further
powers of $1/M_{(4+n)}$ and/or $1/f^2$. 

We should also mention
that if translation invariance is {\it explicitly} broken in
the extra dimensions, as in the case where the wall is ``stuck" to
a point in the higher dimensions, the modes $y^a$ corresponding to
the fluctuations of the wall become massive and are irrelevant to
low energy physics.

\section{Lab bounds}
\subsection{Macroscopic gravity}
Given that the gravitational interaction is unchanged over distances
bigger
than the size of the extra dimensions, and that gravity
is only significant on much larger scales, the change in gravity at
distances smaller than $\sim 1$ mm is harmless.
One may wonder about systems where gravity is known to be important, but
where the typical inter-particle separation is smaller than
$\sim 1$ mm, e.g. in the sun. It is clear, however, that all effects due
to
the new gravity beneath $\sim 1$ mm must be suppressed
by powers in the ratio of the size of the new dimensions over the
typical
size $R$ of the gravitating body. The reason is that, if we
divide the body into $\sim 1$ mm balls, these balls have normal
gravitational interactions. Since important gravitational effects
are bulk effects, the only error incurred in splitting the body into
$\sim$
mm sized balls can at most be power suppressed in
$(1$mm$/R)$. For instance,
let us compute for the gravitational self energy per unit mass of a
ball
of radius $R$ and density $\rho$:
\begin{equation}
\frac{E_{grav}}{\rho R^3}  \sim \int_0^{r_n} d^3 r \frac{G_{N(4+n)}
\rho}{r^{(n+1)}} + \int_{r_n}^{R} d^3 r
\frac{G_{N(4)} \rho}{r}
\end{equation}
where
the first integral uses the $(4+n)$ dimensional gravitational potential
and
the second is the usual piece. Now, the usual
piece is dominated by large distances and gives a contribution $\sim
G_{N(4+n)} \rho R^2$. However, for $n=2$, the new contribution
is log divergent and is cutoff off at short distances by the typical
inter-particle separation $r_{min}$ and at long distances by
$R$, and for $n>2$, the new contribution is dominated by short distances
and is cutoff by $r_{min}$. The fractional change in the
gravitational energy due to the new interaction is then
\begin{equation}
\frac{\Delta E_{grav}}{E_{grav}} \sim \frac{G_{N(4+n)}}{G_{N(4)}
r_{min}^{n-2} R^2} \sim \frac{r_n^n}{r_{min}^{n-2} R^2}.
\end{equation}
Note that for $G_{N(4+n)} \sim ($TeV$)^{-(n+2)}$ and for $r_{min}$
larger
than $\sim ($TeV$)^{-1}$, this contribution
is largest for $n=2$, and
\begin{equation}
\frac{\Delta E_{grav}}{E_{grav}} \lsim \left(\frac{1
\mbox{mm}}{R}\right)^2.
\end{equation}
which is completely irrelevant for the sun. The smallest objects for
which
the gravitational self-energy plays any role is the
neutron star which has $R \sim 10$ km, giving an unobservably small
fractional change $\sim 10^{-12}$ in gravitational energy.

\subsection{Mesoscopic gravity}
While the normal Newtonian gravitation is unaffected on distances larger
than
$r_n$, the gravitational attraction between two objects grows much more
quickly $\sim 1/r^{n+2}$ at distances smaller than $r_n$. This is of
course a reflection
of the fact that, in this scenario, gravity ``catches up" with the other
interactions at
$\sim 10^{3}$ GeV rather than at $10^{19}$ GeV.  The flip side of this
is that, even though
gravity is much stronger than before, it is still much weaker than the
other forces at distances
appreciably larger than the weak scale. Consider for instance the ratio
of the new gravitational force
to the electromagnetic force between a proton and an electron a distance
$r$ apart
\begin{equation}
\frac{F_{grav}}{F_{em}} \sim \frac{G_{N(4+n)} m_e m_p}{\alpha r^n} \sim
10^{-7} \left(\frac{10^{-17} \mbox{cm}}{r}\right)^n.
\end{equation}
The smallest value of $r$ where electromagnetic effects are dominant are
atomic sizes $r \sim 10^{-8}$ cm, and even then
for the worst case $n=2$, the above ratio is unobservably small $\sim
10^{-25}$.
Of course on larger distances the electromagnetic interactions are
screened due to average charge neutrality,
while gravity is not. Even here, however, the residual electromagnetic
forces still dominate over
the new gravity.
As an example consider the Van der Waals (VdW) force between
two hydrogen atoms, in their ground state, a distance $r \gg
r_{\mbox{bohr}}$ apart from each other.
This arises due to the dipole-dipole interaction potential, i.e. the
energy of the dipole-moment of atom 1
in the electric field set up by the dipole moment of atom 2:
\begin{equation}
V_{int.} \sim \frac{d_1 d_2}{4 \pi r^3}.
\end{equation}
The first order energy shift due to this interaction vanishes in the
ground state since
the ground state expectation value of each dipole moment vanishes by
rotational invariance.
The second order perturbation then gives the usual VdW $1/r^6$
potential,
\begin{eqnarray}
\Delta V(r) &\sim& \sum_{n,n'} \frac{d^2_{1 0n} d^2_{2 0n'}}{2E_0 - E_n
- E_n'} \frac{1}{16 \pi^2 r^6} \nonumber \\
 &\sim& \alpha \left(\frac{r_{bohr}}{r}\right)^5 \frac{1}{r}.
\end{eqnarray}
The ratio of this  VdW force to the ordinary gravitational attraction
between the hydrogen atoms is
\begin{equation}
\frac{F_{VdW}}{F_{\mbox{ord. grav.}}}  \sim \left(\frac{1
\mbox{mm}}{r}\right)^5,
\end{equation}
and we see that while electrostatic effects are irrelevant for distances
larger than $\sim 1$ mm,
the VdW force dominates over ordinary gravity at sub-mm distances. This
is in fact the central obstacle to
the sub-mm measurements of gravitational strength forces.
Even in our scenario with much stronger gravity, VdW forces dominate
down to atomic scales,
(where the electromagnetic effects are no longer even shielded). For the
case of $n=2$ new dimensions,
the new dimensions open up near the mm scale, and the gravitational
force only increases as $1/r^4$
at smaller distances, which is still overwhelmed by VdW. Already for
$n=3$, the new dimensions open at
$\sim 10^{-7}$ cm and VdW dominates still further.

Of course in the case $n=2$, we expect a switch from $1/r^2$ to $1/r^4$
gravity roughly beneath $r_2 \sim 1$ mm. There are no direct measurements of
gravity at sub-mm distances.
The best current bound on sub-mm $1/r^3$
potentials actually comes from experiments measuring the Casimir
forces at $\sim 5$ microns \cite{Casimir}. Parametrizing the force
between two
objects composed of $N_1,N_2$ nucleons separated by a distance $r$ as
\beq
V(r) = C N_1 N_2 \frac{(10^{-15} \mbox{m})^2}{r^3}
\eeq
the best current bound is $C \lsim 7 \times 10^{-17}$\cite{Casimir}. If we
assume that the only gravitational strength forces beneath $r_2$
is the $6-$d Newtonian potential, this corresponds to
\begin{eqnarray}
C &=& \frac{G_{N(6)} (m_N \times 10^{15} \mbox{m}^{-1})^2}{3} =
\frac{1 \mbox{GeV}^4}{50 M_{(6)}^4} \nonumber \\
&\to& M_{(6)} \gsim 4.5 \mbox{TeV}.
\end{eqnarray}
If we take this indirect bound seriously, then from eqn.(\ref{rn}),
$r_2$ shrinks to $\sim 30$ microns, which is however still well
within the reach of the planned experiments directly measuring gravity
at
sub-mm distances. There may be contributions to the
long-range force beneath $r_2$ beyond those from the KK
excitations of the ordinary graviton, 
which may compensate the gravitational force and the 
and the force at the $\sim 5$
micron distances probed in the Casimir experiments may not be as strong
we have considered, with correspondingly weaker bounds. If the $1/r^4$ force is canceled at short distances,
a sub-leading $1/r^3$ force may remain.
In this case, the transition from $1/r^2 \to 1/r^3$ 
could be observed for $r_2$ as large
as  $\sim.5$mm. It is interesting that this potential could also be
interpreted as Newtonian gravity in 5 space-time dimensions, with
a new dimension opening up at the millimeter scale!

\subsection{``Compositeness" bounds}
We next discuss laboratory bounds. Since we have quantum
gravity
at the TeV scale, in theory above a TeV
will generate higher dimension operators involving SM fields, suppressed
by
powers of $\sim$ TeV. Of course, operators such as
these which lead to proton decay or large flavor-violations in the Kaon
system
must somehow be adequately suppressed as we have discussed in previous
papers\cite{ADD,AADD}. However, the majority of higher dimension operators suppressed
by
$\sim$ TeV are safe.
Their effects can show up either in modifying SM cross-sections (and are
therefore constrained by
``compositeness" searches), or they can give corrections to precisely
measured observables such as the
electron/muon $(g-2)$ factors or the S-parameter.
Since we do not know the exact theory above
a TeV, the coefficients
of these higher dimension operators are unknown, but we will estimate
their
order of magnitude effects
to show that they do not provide significant constraints on the
framework.
We discuss ``compositeness" constraints first.
The strongest
bounds on 4-fermion operators of the form
\beq
{\cal O}_{4-\mbox{fermi}} = \frac{2 \pi^2}{\Lambda^2}(\bar{\psi} \psi)^2
\eeq
are from LEP searches in the lepton sector, which require at most
$\Lambda \gsim 3.5$ TeV. If the this operator is
generated with coefficient $1/m_{grav}^2$, it is safe for $m_{grav}
\gsim 1$
TeV.

While most of these operators have unknown coefficients, some have
contributions from physics beneath the scale $m_{grav}$ which are in
principle calculable.
For instance, the tree-level exchange of the $(4+n)$ dimensional
gravitons can give rise to local
$4-$fermion operators\cite{AADD}. We can understand this from the 4-dimensional
viewpoint as
follows. If the typical external energy for the
fermions is $\sim E$, then the exchange of a KK excitation of the
graviton
labeled by momenta $(k_1, \cdots, k_n) r_n^{-1}$ with mass $|k|r_n^{-1}
\gsim E$
generates a local 4-fermion operator. Summing over the KK modes yields
an operator of the form
\beq
{\cal O} = C \sum_{|k|r_n^{-1} \gsim E} \times \frac{E^2}{M_{(4)}^2}
\times \frac{1}{|k r_n^{-1}|^2} \times (\bar{\psi} \psi)^2
\eeq

where $C$ is an $O(1)$ coefficient to be
determined
by an exact computation.
For $n=2$, the sum over KK modes is log-divergent in the UV, while for
$n>2$ it is power divergent. Of course, this sum must be cutoff
for the KK modes heavier than $m_{grav}$, where new physics sets in.
For $n=2$, the logarithm is not large enough to significantly
enhance the operator; however, for $n>2$, the power divergence changes
the
$1/M_{(4)^2}$ suppression to an $E^2/m_{grav}^4$ effect:
\beq
{\cal O} = C
\left(\frac{m_{grav}}{M_{(4+n)}}\right)^{n+2} \times
\frac{E^2}{m_{grav}^4} (\bar{\psi} \psi)^2.
\eeq
Of course the precise bound on $m_{grav}$ depends on the relationship
between
$m_{grav}$ and $M_{(4+n)}$. If we take the string scenario and identify
$m_{grav}$ with $m_s$, then this relationship is given in
eqn.(\ref{msm}).
Even in the worst case where the the ``small" radii are not larger
than the string scale $(r_{6-n} m_s = 1$, the bound on $m_s$ coming from
equating the
coefficient of the four-fermi operator with $2 \pi^2/\Lambda^2$ yields
\begin{equation}
m_s > \left(\frac{C g_4^4}{4 \pi^3}\right)^{1/4}
\sqrt{\Lambda E}.
\end{equation}
Since the strongest bounds on $\Lambda$ come from LEP where the energy
is at most
$100$ GeV, we are safe for all $n$ as long as $m_s \gsim$ TeV.

\subsection{Cosmic rays}
While colliders have not yet attained the energies required to probe new
strong quantum
gravitational effects at the TeV scale,  one can wonder whether very
high energy cosmic rays
place any sort of bounds on our scenario.
Indeed, there are very high energy cosmic rays (nucleons) of energies up
to $\sim 10^{20}$ eV =
$10^{8}$ TeV,  eight orders of magnitude more energetic than the
fundamental Planck scale.
Furthermore, when these nucleons impinge on a stationary nucleon, the
center of mass energy
can be as high as $\sim 1000$ TeV.  This raises two questions. First, is
there anything wrong with having
a particle with energy so much larger than the fundamental Planck scale?
And second, do interactions
with such high energies probe post-Planckian physics? The answer
to both questions is no, and we
address them in turn.

It is obvious that there is nothing wrong with having a particle of
arbitrarily high energy,
since energy is not Lorentz invariant. The question is
however, whether a particle can be accelerated
from rest to a Post-Planckian energy. There is certainly no problem
with accelerating a particle to post-TeV energies, as long as the
acceleration is sufficiently small
(but over large enough distances) so that energy loss to ordinary
radiation is negligible.
Note that relevant acceleration scales will be so much smaller than the
weak scale that the couplings to ordinary
radiation vastly dominate the coupling to higher dimension gravitons, so
that as long as ordinary radiation is negligible,
the gravitational radiation energy loss is even smaller.  It is
interesting to note that, in the context of normal gravitational
theory, there have been speculations that it may be impossible to
accelerate a particle to post-Planckian energies; at least
many acceleration mechanisms fail for a variety of reasons
\cite{Nussinov}. As a typical example, suppose that
the acceleration is provided by a constant electric field $E$ acting
over a region of
size $R$. In order to accelerate a charge
$e$ to energy ${\cal E}$, we must have $eER \sim {\cal E}$. On the other
hand, there is an energy $V \sim E^2 R^3$
stored inside the region, which would give a black hole of event horizon
size $R_{\mbox{hor}} \sim V/M_Pl^2=
 ({\cal E}/M_{Pl})^2 R$.
For ${\cal E} \gg M_{Pl}$, the horizon size is much larger than
$R$ and the system would collapse into a black hole.
These sorts of arguments have led to speculations
that perhaps for reasons related to fundamental short-distance physics,
post-Planckian energies are inaccessible.
Our example suggests otherwise:
while may be difficult to accelerate to energies beyond the effective
four dimensional
Planck scale, energies beyond the fundamental short-distance Planck
scale can easily be attained.

Next we turn to the second issue: do cosmic ray collisions with center of mass
energies far above the TeV
significantly probe the physics at distances smaller than $\sim
($TeV$)^{-1}$? The answer to this is obviously no; the
huge fraction of the cross-section for nucleon-nucleon scattering is
diffractive, arising from the finite size of the
nucleon,
giving a typical cross section $\sim 30$ millibarn. The point is of
course that it is not enough for the c.o.m. energy to be large, after
all two particles
traveling in opposite direction with large energies but infinitely far
apart have huge c.o.m. energy but do not interact! In order to probe
short distance physics
at distances $r$, it is necessary to have a momentum transfer $\sim
r^{-1}$; but the vast majority of nucleon-nucleon interactions only
involve
$\sim$GeV momentum transfers. In fact, cosmic rays lose energy in the
atmosphere not through diffractive QCD scattering but
by creating electromagnetic showers, where the effective momentum
transfer per interaction is still smaller.

\subsection{Precision observables}
Corrections to
electron and muon $(g-2)$ are expected to be naturally small
for a very general and well-known
reason. The higher dimension operators which can contribute to e.g. the
electron $(g-2)$ are of the
form
\begin{equation}
{\cal L}_{g-2} \sim \frac{c_5}{m_{grav}}
e^c \sigma^{\mu \nu} F_{\mu \nu} e + \mbox{higher dimensional
operators}.
\end{equation}
Since the lowest dimension operator violates electron chirality, we
parametrize
$c_5 = d_5 m_e/m_{grav}$, and since the QED contribution to $(g-2)$
generates the same operator with
coefficient $\alpha/(\pi m_e)$, the fractional change in $(g-2)$ is of
order
\begin{equation}
\frac{\delta(g-2)}{g-2} \sim d_5 \frac{\pi}{\alpha}
\left(\frac{m_e}{m_{grav}}\right)^2
\end{equation}
which even for $d_5 \sim 1$ and $m_{grav} \sim$ 1 TeV is $\sim
10^{-10}$, smaller than the
experimental uncertainty $\sim 10^{-8}$. The contribution to the muon
$(g-2)$
is similarly safe.
Of course, there are contributions to $d_5$ which can be computed
in the low energy theory involving loops of the light $(4+n)$
dimensional
graviton, in which case $d_5$
is further suppressed by a loop factor,
and the fractional change
in $(g-2)$
is correspondingly smaller.
Furthermore, since all other operators have higher dimension, they will
at most
make comparable contribution to $(g - 2)$. Note that the chirality
suppression of
the dimension $5$ operator was crucial: a $c_5 \sim 1$ is grossly
excluded.
The correct estimate given above indicates why the anomalous magnetic
moment
measurements, in spite of their high precision, do not significantly
constrain
new weak scale physics.

Similar arguments apply to the corrections to precision electroweak
observables.
Consider the graviton loop correction to the $S$ parameter. Again
from the $4-$d viewpoint, we are summing over the contributions of
the towers of KK gravitons. We consider contributions from modes
heavier and lighter than $m_Z$ respectively.
Recall that each KK mode has $1/M_{(4)}$ suppressed couplings. For
the modes lighter than $m_Z$, each contributes $\sim (m_Z/M_{(4)})^2$ to
S. We therefore estimate
\begin{eqnarray}
S_{m_{KK} < m_Z} &\sim& (m_Z/M_{(4)})^2  \times
(m_Z r_n)^n \nonumber \\
&\sim& (m_Z/M_{(4+n)})^{n+2}
\end{eqnarray}
which is a tiny $\lsim 10^{-3}-10^{-4}$ contribution even for the
worst case $n=2$, $M_{(4)}=1$ TeV. For the contribution from a KK mode heavier
than $m_Z$, S also vanishes in the limit $m_{KK} \to \infty$, so the contribution to
$S$ from each mode is $\sim m_Z^4/(M_{(4)}^2 m_{KK}^2)$.
Therefore, the contribution to $S$ from these states is
\beq
S_{m_{KK} > m_Z} = \sum_{|k| r_n^{-1} > m_Z}
\frac{m_Z^4}{M_{(4)}^2 (|k| r_n^{-1})^2}.
\eeq
This is precisely the same sum as was encountered in the
compositeness section, and it is power divergent in the UV for
$n \geq 3$. Cutting the power divergence off at $m_{grav}$, we
find
\beq
S_{m_{KK} > m_Z} \sim (\frac{m_Z}{m_{grav}})^4
\eeq
which even for $n=6$ is $\lsim 10^{-4}$ for $m_{grav} \sim 1$ TeV.

\subsection{Rare decays to higher dimension gravitons}
A far more important set of constraints follow from the fact that the
$(4+n)$ dimensional graviton is a massless particle with
couplings to SM fields suppressed by powers of $\sim 1/$TeV. In this
respect, it is similar to other light particles like axions or
familons. These are known to be in disastrous conflict with experiment
for
decay constants in the $\sim$ TeV region,
for familons because they give rise to large rates for rare
flavor-changing
processes, for axions because they can
take away too much energy from stellar objects through their
copious production. We must check that the
analogous processes do not rule out a $(4+n)$
dimensional graviton with $1/$TeV-suppressed couplings. Another way of
stating the problem is as follows. As we have remarked several times,
from the 4-dimensional
point of view, the graviton spectrum consists of the ordinary
massless graviton, together with its tower of KK excitations
spaced by $r_n^{-1}$. While the coupling of each of
these KK modes is suppressed by $1/M_{(4)}$, there is an enormous number
$\sim (Er_n)^n$ of them available with mass lower than
energy $E$, and there combined effects are much stronger than
suppressions
of $\sim 1/M_Pl$.
This large multiplicity factor is responsible for converting $1/M_{(4)}$
effects
to stronger $/M_{(4+n)}$ effects, as we have already seen explicitly in
the conversion between $1/r^2$ to $1/r^{n+2}$ Newtonian force law.
However, as we have mentioned, the infrared softness of higher dimension
gravity will allow this scenario to survive.
We begin with bounds from rare decays of SM particles involving the
emission of gravitons into the
extra dimensions, beginning with the decay $K \to \pi + $ graviton (the
analogous familon process
$K \to \pi + $ familon puts the strongest bound on familon decay
constants
$\sim 10^{12}$ GeV).
Recall that even though the emission of a single graviton into the extra
dimensions violates conservation
of extra-dimensional momentum,
it is nevertheless allowed, since the presence of the wall on which SM
fields is localised breaks translational invariance in the
extra dimensions. However, since time translational invariance is still
good, energy must still be conserved.
Notice also  that this process will proceed through e.g. the spin-0
component of the massive KK
excitations of the graviton in order to conserve angular momentum.
A tree-level diagram for the process can be obtained by attaching a
$(4+n)$
dimensional graviton to any of the legs of the Fermi interaction
$\bar{s} d
\bar{u} d$. Again, on dimensional grounds, the
decay width for the decay into any single $KK$ mode is at most
\beq
\Gamma_{KK} \sim \left(\frac{1}{16 \pi} \frac{m_K^5}{M_W^4} \right)
\times \frac{m_K^2}{M_{(4)}^2}
\eeq
where the first factor has been isolated as roughly the total KK decay
width.
However, there is a large multiplicity
factor from the number of $KK$ modes with mass $\lsim m_K$ which are
energetically allowed, $\sim (m_K r_n)^n$.
The total width to gravitons is then
\beq
\Gamma_{K \to \pi + \mbox{graviton}} \sim \left(\frac{1}{16 \pi}
\frac{m_K^5}{M_W^4} \right) \times
\left(\frac{m_K}{M_{4+n}}\right)^{n+2}
\eeq
yielding a branching ratio
\begin{equation}
B(K \to \pi + \mbox{graviton})
\sim (m_K/M_{(4+n)})^{n+2}
\end{equation}
Even in the most dangerous case $n=2, M_{(6)} \sim 1 TeV$, this
branching ratio is $\sim 10^{-12}$ and is safely smaller than the bound,
although a more careful calculation is required for
this case. As we will see in the next sections, astrophysics and
cosmology
seem to require $M_{(6)} \gsim 10$ TeV for $n=2$, in which case
the branching ratio in Kaon decay goes down another four orders of
magnitude to $\sim 10^{-16}$.
Note that the scaling for the branching ratio could have also been
derived
directly from the $(4+n)$ dimensional point of view.
As we have remarked earlier, the couplings of the graviton are dimensionless when expressed in terms
of
the $(4+n)$ dimensional metric $G_{MN}$, which can be
expanded about flat space-time as
\begin{equation}
G_{MN} = \eta_{MN} + \frac{h_{MN}}{M_{(4+n)}^{(n+2)/2}}
\label{metric}
\end{equation}
where $h_{AB}$ is the canonically normalized field (of mass dimension $1
+
n/2$) in $(4+n)$ dimensions. Therefore, there is a
factor of $1/M_{(4+n)}^{(n+2)/2}$ in the amplitude and
$1/M_{(4+n)}^{(n+2)}$ in the rate. Inserting factors of the only other
scale, $m_K$ to make a dimensionless branching ratio, we arrive at the
same
estimate for $B(K \to \pi +$ graviton).
We see explicitly that it is the infrared softness of the
interactions of the higher-dimensional theory which is responsible
for insuring safety, although this was certainly not guaranteed
for relatively low $n$.

Analogous branching fractions for flavor-conserving and violating
decays for $B$ quarks are also safe, with branching ratios $\sim
10^{-8}$ for the worst case $n=2, M_{(6)}= 1$TeV, and further down
to $\sim 10^{-12}$ for the $M_{(6)} \gsim 10$TeV favored by
astrophysics and cosmology. The largest branching fractions are
for the heaviest particles, the most interesting being for $Z$
decays. The decay $Z \to f \bar{f} +$ graviton can occur at
tree-level, with a branching fraction
\beq
B(Z \to f \bar{f} + \mbox{graviton}) \sim
(\frac{m_Z}{M_{4+n}})^{n+2}
\eeq
which can be as large as $\sim 10^{-4}$, still not excluded by
$Z$-pole data. Other decays like $Z \to \gamma + $ graviton are
only generated at loop level, with unobservably small branching
ratios.

\section{Astrophysics}
We now turn to astrophysical constraints on our scenario,
analogous to bounds on the interaction of other light particles
such as axions. In our case, the worry is that, since the gravitons
are quite strongly ($\sim 1/$) TeV coupled, they are produced
copiously and escape into the extra dimensions, carrying away energy.
Having escaped, the gravitons have a very small probability to
return and impact with the wall fields: this is intuitively
obvious since the wall only occupies a tiny region of the extra
dimensions. We can also understand this from the point of view of
producing graviton KK excitations. As usual, even though each KK
mode is $1/M_{(4)}$ coupled, significant energy can be dumped into
the KK gravitons because of their large multiplicity. However,
each single KK mode, once produced, has only its $1/M_{(4)}$
coupled interactions with wall fields.
In the next
section we will quantify this correspondence, finding that
the higher dimensional gravitons have a mean-free time for
interaction with wall exceeding the age of the universe for all
graviton energies relevant here. the upshot is that the gravitons
carry away energy without returning energy, thereby modifying
stellar dynamics in an unacceptable way.
For the axion, the strongest such bounds come from SN1987A, which
constrain the axion decay constant $f_a \gsim 10^{9}$ GeV.
This naively spells doom for our $1/$ TeV coupled gravitons.
However, since the gravitons propagate in extra dimensions and
have interactions that are softer in the infrared, our scenario
survives the astrophysical constraints.

We will do a more detailed
analysis below; however, in order to get an idea of what is
going on we estabilsh a rough dictionary between rates for axion
and graviton emission.
Since any axion vertex is suppressed by $1/f_a$, any rate for
axion emission is proportional to
\beq
\mbox{Rate of axion prod.} \propto \frac{1}{f_{a}^2}
\eeq
Now consider graviton production. The
first point is that if the temperature $T$of the star is much smaller
than $r_n^{-1}$, none of the KK excitations of the graviton can be
produced and the only energy loss is the miniscule one to the
ordinary graviton. If $T \gg r_n^{-1}$, on the other hand,
a very large number $\sim (Tr_n)^n$ of KK modes can be produced.
Since each of these modes has couplings suppressed by $1/M_{(4)}$,
the rate for graviton production goes like
\beq
\mbox{Rate of graviton prod.} \propto \frac{1}{M_{(4)}^2} \times (T
r_n)^n \sim \frac{T^n}{M_{(4+n)}^{n+2}}
\eeq
Note that this is exactly analogous to what happened with e.g.
$K \to \pi +$ graviton, and that this dependence
could have been inferred directly from the $(4+n)$-dimensional viewpoint
just as in eqn.(\ref{metric}). We can now establish the rough
dictionary between $f_a$ and $M_{(4+n)}$:
\beq
\frac{1}{f_a^2} \to \frac{T^{n}}{M_{(4+n)}^{n+2}}.
\eeq
This dictionary contains the essence of what will be found by more
detailed analysis below. The strongest bounds come from the hottest
systems (where the bounds on $f_a$ are also the strongest).
However, even for the SN where the average kinetic energy
corresponds $T \sim 30$ MeV, $f_a \gsim 10^{9}$ GeV requires that
for $n=2$ that $M_{(6)} \gsim 10$ TeV, whereas already for $n \geq
3$, $M_{(4+n)}$ can be $\lsim 1$ TeV. Recall also that an $M_{(6)} \sim
10$ TeV can be consistent with new physics (for instance string excitations)
at the $\sim 1$ TeV scale and is therefore not unnatural as far as
the gauge hierarchy is concerned. The constraints from other
systems such as the Sun (where $T \sim$ KeV or Red giants (where $T \sim
100$ KeV)
are weaker and are satisfied for $M_{(4+n)} \lsim 1$ TeV.

We now move to a somewhat more detailed analysis. This is
necessary because there are some qualitative differences between
the axion and graviton couplings;
for instance, the axion coupling to
photons is suppressed not only by $1/f_a$ but also by an ``anomaly
factor" $\alpha/4
\pi$, while there is no corresponding anomaly price for gravitons.
Furthermore, there are some
effects that can not be determined from dimensional analysis
alone, for instance, in some systems, most of the gravitational
radiation comes from non-relativistic particles, and the energy
emission rate depends on the small ratio $\beta = v/c$ in a way
that can not be fixed by dimensional analysis.
It is easy to deduce the dependence on $\beta$ from the couplings to the
physical gravitons. A non-relativistic particle of mass $m$, moving with
some velocity
$\beta_i$ has an energy momentum tensor $T_{\mu \nu} = m (dx_\mu/d
\tau)(dx_\nu/d \tau)$
which in the non-relativistic limit $\beta_i \ll 1$ becomes
\beq
T_{00} = m, \, T_{0i} = p_i, \, T_{ij} = \frac{p_i p_j}{m}.
\eeq
Therefore, the coupling to the physical graviton polarizations,
which come from the transverse,traceless components $h^{ij}$,
has a factor $\sim p^2/m \sim T$ in the amplitude. Therefore,
there is no dependence on $\beta$ from the gravitational vertex.
The situation is different for couplings to
photons; there the fundamental coupling is $e A^{\mu} (dx_\mu/d\tau)$,
and in the non-relativistic limit the coupling to the
physical photons (the transverse part of $A_i$) is suppressed by
$\beta_i$. Of course, for relativistic particles $\beta \sim 1$
and dimensional analysis is all that is needed to estimate the
relevant cross sections.

Since we are concerned with the energy lost to gravitons escaping
into the extra dimensions, it is convenient and standard
\cite{Kolb} to define the quantities $\dot{\epsilon}_{a+b \to
c+grav}$
which are the rate at which energy is lost to gravitons via the
process $a + b \to c + graviton$, per unit time per unit mass of
the stellar object. In terms of the cross-section $\sigma_{a+b \to c+
grav}$
the number densities  $n_{a,b}$ for a,b and the mass density
$\rho$, $\dot{\epsilon}$ is given by
\beq
\dot{\epsilon}_{a + b \to c + grav.} = \frac{\langle n_a n_b
\sigma_{a+b \to c+ grav} v_{rel} E_{grav} \rangle}{\rho}
\eeq
where the brackets indicate thermal averaging.
We can estimate the cross-sections for all graviton
production processes as follows. From the graviton vertex alone,
we get the usual $T^n/M_{(4+n)}^{n+2}$ dependence which already
has the correct dimensions for a cross-section. The dependence on
dimensionless gauge couplings etc. are trivially obtained, while
the appropriate factors of $\beta$ for non-relativistic particles
are dealt with as in the previous paragraph. Finally, we insert an
overall factor $ \delta \sim 1/16 \pi$ to approximately account
for the phase-space.
The relevant processes and estimated cross-sections are shown
below

$\bullet$ Gravi-Compton scattering: $\gamma + e \to e + grav$
\beq
\sigma v \sim \delta e^2 \frac{T^n}{M_{(4+n)}^{n+2}} \beta^2
\eeq

$\bullet$ Gravi-brehmstrahlung: Electron- Z nucleus scattering
radiating a graviton
$(e + Z \to e + Z + grav)$
\beq
\sigma v \sim \delta Z^2 e^2 \frac{T^n}{M_{(4+n)}^{n_2}}
\eeq

$\bullet$ Graviton production in photon fusion: $\gamma + \gamma \to
grav$
\beq
\sigma v \sim \delta \frac{T^n}{M_{(4+n)}^{n+2}}
\eeq

$\bullet$ Gravi-Primakoff process: $\gamma$ + EM field of nucleus Z
$\to grav$
\beq
\sigma v \sim \delta Z^2 \frac{T^n}{M_{(4+n)}^{n+2}}
\eeq

$\bullet$ Nucleon-Nucleon Brehmstrahlung: $N + N \to N + N + grav$
(relevant for
the SN1987A where the
temperature is comparable to $m_\pi$ and so the strong interaction
between N's is unsuppressed)
\beq
\sigma v \sim (\mbox{30 millibarn}) \times
(\frac{T}{M_{(4+n)}})^{n+2}
\eeq

Armed with these cross-sections, we can proceed to discuss the
energy-loss problems in the Sun, Red Giants and SN1987A.

\subsection{Sun}
The temperature of the sun is $\sim 1$ KeV, and the relevant
particles in equilibrium are electron, protons and photons.
The number densities $n_e=n_p$ and $n_{\gamma}$ are roughly
comparable, $\sim n_{e,p,\gamma} \sim($ Kev$)^3$. The electrons and
protons are non-relativistic.
The observed rate at which the sun releases energy per
unit mass per unit time is
\beq
\dot{\epsilon}_{normal} \sim 1 \mbox{erg g}^{-1} \mbox{s} ^{-1} \sim
10^{-45} \mbox{TeV}.
\eeq
We must therefore demand that the rate of energy loss to gravitons
is less than this normal rate.
We will consider the processes in turn.
Begin with the Gravi-Compton scattering.
Using $n_e/\rho = n_p/\rho = 1/m_p$
and $n_\gamma \sim T_{sun}^3$, we find
\beq
\dot{\epsilon} \sim 4 \pi \alpha \delta \frac{T_{sun}^{n+5}}{m_p m_e
M_{(4+n)}^{n+2}}
\eeq
and therefore
\beq
M_{(4+n)} \gsim 10^{\frac{16 - 6n}{n+2}} \mbox{GeV}.
\eeq
Even the worst case $n=2$ only requires $M_{(4+n)} \gsim 10$
GeV.
Gravi-brehmstrahlung is not relevant  since there are
no high-Z nuclei present in the sun.
Photon pair fusion into graviton is more important than the the
analogous process
$\gamma + \gamma \to$ axion, which is highly suppressed by the
``anomaly price" $\alpha/4 \pi$. For the case of graviton, the
rate is given by
\beq
\dot{\epsilon} \sim \delta \frac{T_{sun}^{n+7}}{\rho
M_{(4+n)}^{n+2}}.
\eeq
This places a lower bound on $M_{(4+n)}$,
\beq
M_{(4+n)} \gsim 10^{\frac{18 - 6n}{n+2}}GeV.
\eeq
For $n=2$, this is a stronger bound $M_{(6)} \gsim 30$ GeV, but
certainly no problem.

The Gravi-Primakoff (with photons scattering off the electric
field of the protons) is sub-dominant to the last bound because,
while protons and photons have roughly equal number density, the
electric field surrounding a proton is proportional to the
electric charge $e/4 \pi$ and so the Gravi-Primakoff rate is suppressed
relative to the photon-photon fusion by rate by $\sim \alpha$.

Finally, nucleon-nucleon brehmstrahlung is is irrelevant because
at these temperatures, the collisions of nucleons can not probe
the strong interaction core.

It is clear that the situation with the sun is so safe because
it's temperature is so low. Because electrons, protons and
photons occur in equal abundance, but the cross-sections involving
photons and electrons are suppressed by $\alpha$ and $\beta^2$
effects, the dominant process is the photon-photon fusion, which
yields even for the worst case $n=2$, $M_{(6)} \gsim 30$ GeV.
For red giants, the temperature is somewhat larger, $T \sim 10$
KeV, and the constraints are somewhat different, but the
temperature is still so low that certainly $M_{(4+n)} \sim 1$ TeV
is safe for all $n$.
Clearly the strongest bounds will come
from SN1987A where the temperature is significantly higher $\sim 30$
MeV. We turn there now.

\subsection{SN1987A}
During the collapse of the iron core of SN1987A, about $10^{53}$ ergs of
gravitational binding energy was released in a few seconds; the
resulting neutron star had a core temperature $\sim 30$ MeV. We
must ensure that the graviton luminosity does not exceed the
liberated $10^{53}$ erg s$^{-1}$:
\beq
{\cal L}_{grav} = \dot{\epsilon} M_{SN} \lsim 10^{53} erg
\mbox{s}^{-1} \sim (10^{16} \mbox{GeV})^2
\eeq

There are two dominant processes here: nucleon-nucleon
brehmstrahlung (which is the dominant process for axions), together
with the Gravi-Primakoff process (which is again sub-dominant in
the axion case because the ``anomaly factor" $\alpha/4 \pi$).
The graviton luminosity from the nucleon-nucleon
brehmstrahlung is roughly
\beq
{\cal L}_{grav} \sim M_{SN} \times \frac{n_N^2}{\rho} \times 30
\mbox{millibarn}
\times \left(\frac{T}{M_{4+n)}}\right)^{n+2}.
\eeq
For $M_{SN} \sim 1.6 M_{sun} \sim 10^{57}$ GeV, $n_N \sim 10^{-3}$
GeV$^3$ and $\rho \sim 10^{-3}$ GeV$^4$, we find the following
bound on $M_{(4+n)}$
\beq
M_{(4+n)} \sim 10^{\frac{15 - 4.5n}{n+2}} \mbox{TeV}.
\eeq
For $n=2$, this is quite a strong bound, requiring $M_{(6)} \gsim 30$
TeV.
We next estimate the graviton luminosity from the
Gravi-Primakoff process.
Using $n_{Fe}/\rho \sim 1/m_{Fe}$ and $Z \sim
50$, we have
\beq
{\cal L}_{grav} \sim 10^{57}\mbox{GeV} \delta Z
\frac{T_{SN}^{n+4}}{M_{(4+n)}^{n+2}}
\eeq
which requires
\beq
M_{(4+n)}  \gsim 10^{\frac{12 - 4.5 n}{n+2}} \mbox{TeV}.
\eeq
This is a somewhat weaker bound than for nucleon-nucleon brehmstrahlung.
The basic reason is that while again in the SN, nucleon and photon
abundances are comparable (actually nucleons are somewhat more
abundant), the nucleon-nucleon brehmstrahlung cross-section is
enhanced by strong-interaction effects.

In summary, we have found as expected that the strongest
astrophysical bounds come from the hottest system, SN1987A.
The bounds for $n=2$ were quite strong, requiring $M_{(6)} \gsim 30$
TeV. This illustrates that the phenomenological viability of our
scenario is not an immediate consequence of localizing the SM
particles on a wall. Nevertheless, for $n>2$, the infrared
softness of higher dimensional gravity was enough to evade the
constraints for $M_{(4+n)} \sim 1$ TeV. Even for $n=2$, 
$M_{(6)} \sim 30$
TeV is consistent with a string scale $\sim few$ TeV, and
therefore this case is still viable for solving the hierarchy problem
and
accessible to being tested at
the LHC.

\section{Cosmology}
It is clear that in our scenario, early universe cosmology is
drastically different than the current picture. Since the
fundamental short distance scale is $\sim$ 1 TeV, the highest
temperature at which we can conceivably think about a reasonable
space-time where the universe is born is $\sim m_{grav} \sim TeV$
rather than $M_{(4)} \sim 10^{19}$ GeV. Even beneath these
temperatures, however, the dynamics of the extra dimensions is
critical to the behavior of the universe on the wall. In the
absence of any concrete mechanism for stabilizing the radius of
the extra dimensions, we can not track the history of the universe
starting from TeV temperatures. Of course, nothing is known
directly about the universe at TeV temperatures. The only aspect
of the early universe which we know about with some certainty
is the era of Big-Bang Nucleosynthesis (BBN) which begins at
temperatures $\sim 1$ MeV.
The successful predictions of the light
element abundances from BBN implies that the expansion rate of the
universe during BBN can not be modified by more than $\sim 10 \%$.
Since the size of the extra dimensions determines $G_{N(4)}$
and hence the expansion rate of the $4-d$ universe on the wall, we
know that whatever the mechanism for stabilizing the
extra-dimensional radii, they must have settled to their
current size before the onset of BBN. Note that the
radii must be fixed with size $\lsim$ mm, which is much smaller than the
Hubble size $\sim 10^{10}$ cm at
BBN. Therefore, the expansion of the 4-d universe can be described
by the usual 4-d Robertson-Walker metric. This is analogous to the
analysis of macroscopic gravity in section 5.1, where we saw
that even when inter-particle separations are smaller than $r_n$, the
large-distance gravitational energetics are unaffected.
Furthermore, the
extra-dimensions must be relatively empty of energy-density, since
this would also contribute to the expansion rate of the 4-d
universe.

This leads us to parametrize our
ignorance about the physics determining the radius as follows.
Extrapolating back in time from BBN, we assume that
the universe is ``normal" from BBN up to some maximum temperature
$T_*$ for the wall states. By ``normal" we mean that
the extra dimensions are essentially frozen and empty
of energy density. One possible way this initial condition can
come about is if $T_*$ is the re-heating temperature after a period
of inflation on the wall. The inflaton is a field localised on the
wall and its decays re-heat predominantly wall-states while not
producing significant numbers of gravitons.

We will test the consistency and
cosmological viability of such a starting point.
The main reason this will be non-trivial is due again to the
presence of light modes other than SM particles- namely the
extra-dimensional gravitons and, for the case where the wall is
free to move, the goldstones describing the position of the wall.

It is easy to see that the goldstones are not especially
problematic: they have a very small mass $\sim 10^{-3} eV$,
and since they are their own antiparticles, they would
count as $n/2$ extra neutrinos during Nucleosynthesis
if they have thermal abundance. For $n=2$,
this is marginally consistent with BBN, whereas for $n>2$
we have to insure that they are not thermal during BBN.
This puts some upper bound on the ``normalcy" temperature $T_*$.
If the (model-dependent) coupling $\sim \lambda \psi \partial^{\mu}
\psi^c \partial_\mu g/f^2$
is responsible for thermalization, the goldstone drops out of
equilibrium when
\begin{eqnarray}
T_* &\lsim& \frac{f^{4/3}}{M_{Pl}^{1/3} \lambda^{2/3}_{max}(T_*)}
\nonumber \\
& \to & \lambda^{2/3}(T_*) T_* \lsim 10^{-2} \mbox{GeV}
\end{eqnarray}
where $\lambda_{max}(T_*)$ is the largest Yukawa coupling of a
SM particles thermal at temperature $T_*$.This roughly translates
to $T_* \lsim 1$ GeV. If instead the model-independent couplings
suppressed by $1/f^4$ are keeping equilibrium, decoupling happens
when
\beq
T_* \lsim \frac{f^{8/7}}{M_{Pl}^{1/7}} \sim 10 \mbox{GeV}.
\eeq
This is a weak bound for obvious reasons:
the goldstones are essentially
massless, with smaller interaction cross-sections than neutrinos,
and so it is guaranteed that they decouple before BBN, where neutrinos
decouple.
Furthermore, since they are so light, these goldstones can not
over-close the universe.

Gravitons provide further cosmological challenges.

$\bullet$ {\bf Expansion dominated cooling}

The energy density of the radiation on the wall cools in two ways. The
first is the normal cooling
due to the expansion of the universe:
\beq
\frac{d \rho}{d t}|_{\mbox{expansion}} \sim - 3 H \rho \sim -3
\frac{T^2}{M_{Pl}} \rho
\eeq
 The second is cooling by
``evaporation" into the extra dimensions, by  producing  gravitons which
escape into the
bulk.
Notice again that this sort of cooling does not occur if the SM
fields couple to some generic $1/$
TeV coupled but 4-dimensional particle X, since the rates for the
forward and
backward reactions would proceed at the same rate and X would
thermalize. The rate for graviton production is proportional to
the usual factor $1/M_{(4+n)}^{n+2}$, and the rate for
evaporative cooling can be determined by dimensional analysis to
be
\beq
\frac{d \rho}{d t}|_{\mbox{evap.}} \sim - \frac{T^{n +
7}}{M_{(4+n)}^{n+2}}
\eeq
The expansion rate of the universe can only be normal if the
rate for normal expansion by cooling is greater than the that from
evaporation. This put an upper bound on the temperature $T_*$ at
where the universe can be thought of as normal:
\beq
T_* \lsim \left(\frac{M_{(4+n)}^{n+2}}{M_Pl}\right)^{1/(n+1)} \sim
10^{\frac{(6 n - 9)}{(n+1)}} \mbox{MeV} \times \left(\frac{M_{(4+n)}}{1
\mbox{TeV}}\right)^{(n+2)/(n+1)}
\eeq
For the worst case $n=2$, this is $T_* \lsim 10$ MeV for $M_{Pl(4+n)}
\sim 1$
TeV. However, the astrophysical constraints prefer $M_{(4+n)} \sim 10$
TeV, in which case $T_*$ moves up to $\lsim 100$ MeV, while for
$n=6$, $T_* \lsim 10$ GeV. Of course,as $n \to \infty, T_* \to
M_{(4+n)}$. It is reassuring that in all cases, $T_* \gsim 1$ MeV,
so that BBN will not be significantly perturbed.

We can understand this constraint in another way. The rate of
production of $(4+n)$ dimensional gravitons produced per relativistic
species (``photons") on the wall, is given by
\beq
\frac{d}{dt} \frac{n_{grav}}{n_{\gamma}} = \langle n_{\gamma}
\sigma_{\gamma \gamma \to grav} v \rangle \sim
\frac{T^{n+3}}{M_{(4+n)}^{n+2}}
\eeq
so that the total number density of gravitons produced during a
Hubble time starting at temperature $T_*$ is
\beq
\frac{n_{grav}}{n_{\gamma}} \sim \frac{T_*^{n+1}
M_{Pl}}{M_{(4+n)}^{n+2}}
\eeq
The ``cooling" bound we have given corresponds to requiring
$n_{grav} << n_{\gamma}$.

$\bullet$ {\bf BBN constraints}

We must ensure that the produced gravitons do not significantly
affect the expansion rate of the
universe during BBN. The energy density in gravitons red-shifts
away as $R^{-3}$ rather than $R^{-4}$. This is because, from the
4 dimensional  point of view, the gravitons produced at temperature $T$ are
massive KK modes with mass $\sim T$. Alternately, from the
$(4+n)$ dimensional point of view, while the graviton is massless, the extra
radii are frozen and not expanding, so the component of the
graviton momentum in the extra dimensions is not red-shifting.
The ratio of the energy density in gravitons versus photons by the
time of BBN is then
\beq
\frac{\rho_{grav.}}{\rho_{\gamma}}|_{\mbox{BBN}} \sim \frac{T_*}{1
\mbox{MeV}} \times \frac{T_*^{n+1} M_{Pl}}{M_{(4+n)}^{n+2}}
\eeq
Therefore, to insure normal expansion rate during BBN, the bound on $T_*$ is 
slightly
stronger
\beq
T_* \lsim 10^{\frac{6n - 9}{n+2}} \times \frac{M_{(4+n)}}{1
\mbox{TeV}}.
\eeq

$\bullet$ {\bf Over-closure by gravitons}

The constraints we have discussed above would equally well apply
to the production of purely purely 4-d particles with $1/$TeV
suppressed couplings of the appropriate power. The production of-
gravitons is, however,
qualitatively different since they escape into the bulk, with a
very low probability of returning
to interact with the SM fields on the wall.
Consider the width $\Gamma$ for a graviton propagating with energy
$E$ in the bulk, to decay into two photons on the wall. This
interaction can only take place if the graviton is within its
Compton wavelength $\sim E^{-1}$ from the wall. The probability that
this is the case in extra dimensions of volume $r_n^n$ is
\beq
P_{\mbox{grav. near wall}} \sim (Er_n)^{-n}
\eeq
On the other hand, when it is near close to the
wall, it decays into photons with a coupling suppressed by $\sim
M_{(4+n)}^{-(n+2)/2}$, and therefore the width is
\beq
\Gamma_{\mbox{near wall}} \sim \frac{E^{n+3}}{M_{(4+n)})^{n+2}}
\eeq
The total width $\Gamma$ is
\beq
\Gamma = P_{\mbox{grav. near wall}} \times \Gamma_{\mbox{near
wall}} \sim \frac{E^3}{r_n^n M_{(4+n)}^{n+2}} \sim
\frac{E^3}{M_{(4)}^2}
\label{gam}
\eeq
This simple result could have also been understood directly from
the KK point of view: the coupling of any KK mode is suppressed by
$1/M_{(4})$, so the width for any individual KK mode to go into SM
fields is suppressed by $1/M_{(4)}^2$ and the above width follows
from dimensional analysis.
Of course significant
amounts of energy can be lost to these KK modes, despite their
weak coupling, for the usual reason of their enormous multiplicity.
Among other things, eqn.(\ref{gam})
implies that the gravitons can be very long-lived, since they can
not decay in the empty bulk . This is because, as long as
the momenta in the extra dimensions is conserved, the graviton
(which is massless from the (4+n) dimensional point of view)  can not decay into
two other massless particles. Of course, interaction with the wall
breaks translational invariance and allows momentum
non-conservation in the extra dimensions, but this requires that
the decay take place on the wall. The lifetime of a graviton of
energy $E$ is then
\beq
\tau(E) \sim \frac{M_{(4)}^2}{E^3} \sim 10^{10} \mbox{yr} \times
\left(\frac{100 \mbox{MeV}}{E}\right)^3.
\eeq

The gravitons produced at temperatures beneath $\sim 100$ MeV have
lifetimes of at least the present age of the universe.
The ratio $n_{grav}/n_{\gamma}$ which was constrained to be $\lsim 1$
in the above analysis must be in fact  much smaller in order for the
gravitons not to over-close the universe.
As we have mentioned,  most of the gravitons are ``massive" with mass
$\sim T_*$
from the 4-d point, they dramatically over-close the universe
if their abundance is comparable to the photon abundance at early times.

The energy density stored in the gravitons produced at temperature
$T_*$ is
\beq
\rho_{grav} \sim T_* \times n_{grav} \sim \frac{T_*^{n+5}
M_{Pl}}{M_{(4+n)}^{n+2}}
\eeq
which then red-shifts mostly as $R^{-3}$. The ratio
$\rho_{grav}/T^3$ is invariant. The critical density of the universe
today corresponds to
to $ (\rho_{crit}/T^3) \sim 3 \times 10^{-9}$ GeV. For the gravitons not
to over-close the universe, we therefore
require for critical density at the present
age of the universe. We therefore require
\begin{eqnarray}
3 \times 10^{-9} \mbox{GeV} \gsim & \rho_{grav}/T_*^3 & \sim
\frac{T_*^{n+2} M_{Pl}}{M_{(4+n)}^{n+2}} \nonumber \\
\to & T_* \lsim 10^{\frac{6n - 15}{n+2}} \mbox{MeV} \times
\frac{M_{(4+n)}}{\mbox{TeV}}
\end{eqnarray}
This is a serious constraint. For $n=2$, we have to push
$M_{(4+n)}$ to the astrophysically preferred $\sim 10$ TeV,
to even get $T_* \sim 1$ MeV, although of course in this case a
much more careful analysis has to be done. For $n=6$, we need $T_* \lsim
300$
MeV.

$\bullet${\bf Late decays to photons}

Finally, we discuss the bounds coming from the late decay of
gravitons into photons which would show up today as distortions of
the diffuse photon spectrum. For $T_* \lsim 100$ MeV, the graviton
lifetime is longer than the age of the universe by $\sim (100$
MeV$/T_*)^3$
, but a fraction $\sim (T_*/100$ MeV$)^3$ of them have already
have already decayed, producing photons of energy $\sim T_*$. The
flux of these photons (i.e. the number passing through a given
solid angle $d \Omega$ per unit time) is then roughly
\beq
\frac{d {\cal F(T_*)}}{d \Omega} \sim n_{0 grav} H_0^{-1} \times
(\frac{T_*}{\mbox{100 MeV}})^3.
\eeq
This is to be compared with the observational bound on the diffuse
background radiation at photon energy $E$, which can be fit approximately
by
\beq
\frac{d {\cal F(E)}}{d \Omega} \lsim \frac{1 \mbox{MeV}}{E}
\mbox{cm}^{-2} \mbox{sr}^{-1} s^{-1}
\eeq
Using the previously derived expressions for the present
$n_{grav}$, this gives us a bound on $T_*$,
\beq
T_* \lsim 10^{\frac{6 n - 15}{n+5}} \mbox{MeV} \times
\left(\frac{M_{(4+n)}}{\mbox{TeV}}\right)^{\frac{n+2}{n+5}}
\eeq
Again, for $n=2$, even pushing $M_{(4+n)}$ to $\sim 10$ TeV pushes
$T_*$ up to only $\lsim 1$ MeV. On the other hand,
for $n=6$ and $M_{10} \sim 1$ TeV, $T_* \lsim 100$ MeV is safe.

Notice that the bound from photons always demands a $T_*$ which is
lower than that which critically closes the universe. Therefore,
in this minimal scenario, the KK gravitons can not account for the
dark matter of the universe. Of course this is not a problem,
the dark matter can be accounted for by other states in the theory.
Given the inevitability of graviton production, however, graviton
dark matter would certainly be attractive.
There is a way out of the bound from decay to photons which can
make this possible.

$\bullet$ {\bf Fat-branes in the bulk}

The problem arose because we assumed
that, once the graviton is emitted into the extra dimensions, it
must eventually return to our 4-d wall in order to decay.
Suppose however that there was another brane in the bulk, of
perhaps a different dimensionality. Since gravity couples to
everything, it could in particular couple to the matter on
this new wall and lower the branching ratio for decaying on our
wall. In fact, if the new wall has more than three
spatial dimensions, the branching ratio to decay into photons on
our wall would be drastically reduced. This can be seen in a
number of ways. Suppose that the new wall has (3+$p$) spatial dimensions
with
$p \leq n$. Note that since the extra dimensions are compactified,
the extra $p$ spatial dimensions are not infinite but have size $\sim
r_n$. We will call this new wall a  fat-branes. 
Now, a graviton propagating in the bulk with energy $E \gg r_n^{-1}$ 
cannot resolve the difference between this new wall and stacks of $(Er_n)^p$
normal $3-d$ walls spaced $E^{-1}$ apart. But then, the branching
ratio for the graviton to decay on our wall is reduced greatly
by $(E r_n)^{-p}$. The width for gravitons to decay on the new
wall is
\beq
\Gamma \sim \frac{T^3}{M_{Pl}^2} \times (T r_n)^p \sim
\frac{T^{p+3}}{M_{Pl(4+p)}^{p+2}}
\eeq
where we have used the relationship between Planck scales of
different dimensionalities in the final expression. This also
gives another interpretation of the result. From the viewpoint of
a graviton of energy $E \gg r_n^{-1}$, the fat-brane may as well
be infinite in all $3+p$ dimensions.
Therefore, just as the width to
decay on our wall is small because the interaction of
any single graviton KK mode is suppressed by $1/M_{(4)}$, so the
width to decay on the other wall is suppressed by
$M_{(4+p)}^{-(p+2)/2}$. The branching ratio is then bigger
because the higher dimensional Planck scale relevant to the
$(3+p)$-brane is smaller.
The lifetime for the graviton to decay on the fat-brane can easily
be much smaller than the age of the universe.
What is the fate of gravitons which decay on the fat-brane?

$\bullet$ {\bf Dark matter on the fat-brane}

In order to understand the evolution of the universe after the
decay of gravitons on the fat-brane, it is important to understand
the cosmology of the fat-brane itself. There are two important
points. First, just as for our 3-brane, at distances
larger than $r_n$ gravity on the fat-brane is normal and four-dimensional. This is
because on
scales larger than $r_n$, the ``thickness" of the fat-brane can
not be resolved. Second, the energy densities on all branes
contribute to the 4-d expansion rate of both our brane and the
fat-brane. Therefore, there is effectively a single energy density
and a common 4-d expansion rate for the two branes.
Consequently, the way that the expansion rate is affected after
the gravitons are captured on the fat-brane depend on the nature
of the decay products there. If they are non-relativistic, their
energy density red-shifts away like $R^{-3}$ and they may provide a
dark matter candidate. Notice that this dark matter may actually
``shine" on its own brane; it is only dark to us. This allows any mass range for the
dark matter candidates, since they can never into ordinary photons.

\section{TeV Axion in the bulk and the strong CP problem}
As we have remarked, the main reason our scenario remains
phenomenologically viable is that the couplings to states that
can propagate in the bulk are suppressed. This observation can
also be used to revive the TeV axion as a solution to the strong
CP problem, if the axion is taken to be a bulk field.
Without specifying the origin of the axion, the relevant terms in the
low-energy effective theory are
\beq
{\cal L}_{eff} \supset \int d^{4+n} x (\partial a)^2 + \int d^4 x
\frac{a(x;x^a=0)}{f_a^{(n+2)/2}} F \tilde{F}.
\eeq
where $a=4,\cdots,3+n$ runs over the extra dimensions.
Just as always, QCD will generate a potential for $a(x,x^a=0)$. 
In order to minimize energy, $a(x,x^a=0)$ 
will prefer to sit at the minimum of this
potential, solving the
strong CP problem on the wall. Furthermore, in order to minimize kinetic energy, $a$
will take on this vev uniformly everywhere in the bulk. From the 4 dimensional point
of view, we can expand $a$ into KK excitations. After going to canonical
normalization, each of these has $1/M_{(4)}$ suppressed couplings to 
$F \tilde{F}$ for $f \sim M_{(4+n)} \sim$ TeV.
The potential 
that is generated by QCD is then minimized with the zero mode acquiring the
appropriate vev  and all the massive modes having zero vev. 

An explicit field theoretic model producing such an axion
field can be easily constructed. Let $u^c, d^c$ and $Q$ be the weak doublet and singlet
quark fields respectively and $H$ be a electroweak Higgs doublet. In our theory
these states are the
four-dimensional modes on the
3-brane. Let $\chi$ be a bulk complex scalar field whose  
spatially constant vev will break PQ symmetry.
\begin{equation}
\langle \chi \rangle \sim M_{(4+n)}^{1 + {n \over 2}} \label{axive}
\end{equation}
The $4+n$ dimensional axion field is defined as
\beq
a = \langle\chi\rangle arg \chi = \frac{a_4(x_{\mu})}{\sqrt{r_n^n}}
 +
\mbox{KK- modes}
\eeq
where we have expanded into KK modes.
As already mentioned, the zero mode $a_4$ is a genuine
four-dimensional axion field, with the $1/r_n^{{n \over 2}}$ 
insuring that its 4-d kinetic term is canonically normalized. 
The coupling of $\chi$ with matter on the 3-brane 
can be written as
\begin{equation}
 \int d^{n + 4}x \delta(x^a){ \chi \over M_{(4+n)}^{1 + {n \over 2}}}
\left (H Qu^c +
H^{*}Qd^c \right )
\end{equation}
It is straightforward to see that an effective coupling of the genuine axion to
$F \tilde{F}$ is
\begin{equation}
 \sim {a_4 \over \langle \chi \rangle r_n^{n/2}} F\tilde F \sim {a_4 \over M_4} F \tilde F
\end{equation}
and thus from the point of view of the four-dimensional theory 
it is effectively a Planck-scale axion.
While the bulk axion field $a$ has only 1/TeV suppressed couplings,
it is safe from 
all astrophysical constraints we have considered
for the same reason gravitons are safe. 
Of course, 4-d axions with such high decay
constants ordinarily suffer from the usual cosmological moduli
problem \cite{axion}; we have nothing to add to the early cosmology which needs to
drive the axion to the origin. However,as long as the axion is at
its origin at temperature $T_*$, it will not be significantly
excited during the subsequent evolution of the universe, again for
the same reason gravitons were not significantly  excited.

\section{Gauge Fields in the Bulk}

For a variety of reasons, it seems unlikely that 
$SU(3)\otimes SU(2)\otimes U(1)$ is the only gauge group under which 
the SM fields are charged.
Normally, the non-observation of additional gauge particles is attributed to a very
high scale of symmetry breaking $\gsim$ TeV and comparably high masses for the gauge
bosons. The impact of these heavy gauge bosons on low energy physics is then very
limited. By contrast, in this section we will see that the situation can change
dramatically in theories with large extra dimensions. This can happen if 
if the new gauge bosons 
can freely propagate in the bulk \footnote{The gravi-photons
are  
model-independent examples of this sort.}, while matter charged under the gauge
group, including scalars which may spontaneously break the symmetry, live on a
3-brane. 
The following features emerge: independent of the number of extra
dimensions, the gauge field can mediate a repulsive force more than 
a million times stronger
than gravity at distances smaller than a millimeter.
This raises the exciting possibility that these forces will be discovered in
the measurements of sub-mm gravitational strength forces \cite{sub-mm}. 

Consider for simplicity a $U(1)$ gauge field propagating in the bulk. The free
action is
\beq
{\cal L} = \int d^{4+n} x \frac{1}{4 g_{4+n}^2} F^2, 
\, F_{MN} = \partial_M A_N -
\partial_N
A_M
\eeq
where we take the scale of the dimensionful $(4+n)$ dimensional gauge
coupling to
be $\sim$ the ultraviolet cutoff $M_{(4+N)}$:
\beq
g^2_{(4+n)} \sim  M_{(4+n)}^{-n}
\eeq
This gauge field interacts with matter fields living on a 3-brane via the induced
covariant derivative on the brane
\beq
D_{\mu} \phi = (\partial_{\mu} + i q_{\phi} A_\mu(x,x^a=0)) \phi.
\eeq
Expanding $A_{\mu}$ in KK modes, only the zero mode $A^0(x)$  
transforms under wall field gauge transformations 
$\phi \to e^{i \theta(x)} \phi$; the rest of the KK modes are massive starting at 
$r_n^{-1}$. At distances much larger than $r_n^{-1}$, only the zero mode is
relevant, and the action becomes
\beq
{\cal S} = \int d^4 x \frac{1}{g_4^2} (\partial_{\mu} A^0_{\nu} - \partial_{\nu}
A^0_{\mu})^2 + {\cal L_{\mbox{matter}}}(\phi,D^0_{\mu} \phi)
\eeq 
where the effective 4-dimensional gauge coupling is 
\beq
g_4^2 \sim \frac{1}{r_n^2 M_{(4+n)}^n} \sim 
\frac{M_{(4+n)}^2}{M_{(4)}^2}.
\eeq
The first interesting point is that this is a miniscule gauge coupling $g_4 \sim
10^{-16}$ for $M_{(4+n)} \sim 1$ TeV, independent of the number of extra
dimensions $n$. Suppose this gauge
field couples to protons or neutrons. The ratio of the repulsive force mediated 
by this gauge field to the gravitational attraction is 
\beq
\frac{F_{\mbox{gauge}}}{F_{grav}} \sim \frac{g_4^2}{G_N m_p^2} \sim 10^6
\left(\frac{g_4}{10^{-16}}\right)^2  
\eeq
Clearly, the corresponding gauge boson can not remain massless. If the gauge
symmetry is broken by the vev of a field 
$\chi$ on the wall, the gauge
boson will get a
very small mass, which is however exactly in the interesting  
range experimentally:
\beq
m_{A^0}^{-1} = (g_4 q_{\chi} \langle \chi \rangle)^{-1} \sim 1 \mbox{mm} 
\times \left(\frac{10^{-16}}{g_4}\right) \left(\frac{1 \mbox{TeV}}{q_{\chi}
\langle \chi \rangle}\right)
\eeq
Of course there are a number of undetermined parameters so a hard prediction is
difficult Nevertheless, it is reasonable that $g_4 \sim 10^{-16}$ is a lower
bound, and so we can expect repulsive forces between say $10^6-10^8$ times
gravitational strength at sub-mm distances. For all $n>2$, the mass of the 
KK excitations of the gauge field are too large to give a signal at 
the distances probed by the next generation sub-mm force experiments. The case
$n=2$ has still richer possibilities since the KK excitations will have comparable
masses to the lowest mode, and may
contribute significantly to the measured long-range force.

The most interesting possibility is to relate 
this new gauge field with the global Baryon ($B$) or Lepton
($L$) number symmetries of the standard model. 
The gauging of the anomaly-free $B-L$ symmetry
has a definite experimental signal: since atoms are neutral, 
the $B-L$ charge of an atom is its neutron
number. Thus, the hydrogen atom will not feel this force, while 
it will be isotope dependent for other materials.
Gauging other combinations of $B$ and $L$,  e.g. either $B$  or $L$ 
separately, is very interesting as well. Let us consider the case of gauging
baryon number. Of course we have to worry about canceling anomalies; the most
straightforward way out is to add chiral fermions canceling the anomaly which
become massive when $SU(2)_L \times U(1)_Y$ is broken. For instance, we can add
three extra generations with opposite baryon numbers (ignoring the obvious
problem with the $S$-parameter). The interest in this exercise is that it 
may provide a mechanism for suppressing proton decay. Although baryon number
must be broken, dangerous proton decay operators may be tremendously 
suppressed if 
the higgs that breaks $B$ lives on a different brane. 

\section{Cosmological Stability of Large Radii}
We have said nothing about what fixes the radii of the extra
dimensions at their large values, this is an
outstanding problem. The largeness of the extra
begs another, more dramatic question: why
is our 4-dimensional universe so much larger still? It is not
considered a failing of the SM that it offers no explanation of
why the universe is so much larger than the Planck scale. Indeed,
this is equivalent to the cosmological constant problem. If the
density of the universe at the Planck time was $O(1)$ in Planck
units, there would be no other time scale than the Planck time and
the universe would not grow to be $10^{10}$ years old. This was only
possible because the energy density is so miniscule compared to
$M_{Pl}^4$, which is precisely the cosmological constant problem.
It may be that once the cosmological constant problem is
understood, whatever makes the enormity of our 4-d universe
natural can also explain the (much milder) largeness of the extra $n$
dimensions. 

In this context we would like to make the side
remark that the usual cosmological constant problem is in some
sense less severe in our framework. Suppose for instance that
there is a string theory with string tension $m_s \sim $TeV, but
where the SUSY is primordially broken only on our wall at the scale
$m_s$.
As we argued in \cite{AADD}, the SUSY breaking mass splittings
induced for bulk modes is then highly suppressed $\sim (1$ mm$)^{-1}$
at most. However, there is nothing that can be done about the $\sim($TeV$)^4$
vacuum energy on the wall, and we have to imagine canceling it
by fine-tuning it away against a bare cosmological constant
\beq
\int d^{4 + n} x \sqrt{-g} \Lambda_{0(4+n)} \to \int d^4 x \sqrt{-g}
(r_n^n
\Lambda_{0(4+n)}).
\eeq
Since the radii $r_n$ are large compared to  (TeV)$^{-1}$, the
mass scale $\Lambda^{1/(4+n)}_{0(4+n)}$ does not have to be as
large as the TeV scale to cancel the cosmological constant.
Note that since the SUSY splittings in the bulk are so small,
there is no worry of an $\sim$(TeV$)^{4+n}$ cosmological constant
being generated.

We do not, however, have to hide behind our ignorance about the
cosmological constant problem. It may be that the radii are large for
more mundane reasons: for some reason, some of the radius moduli
have a potential energy
with a minimum at very large values of $r m_{grav}$. Even without
knowing anything about the origin on such a potential, we can
place phenomenological constraints on $V(r)$ by requiring that the
field was not significantly perturbed from its minimum by
interacting with the hot universe from the time of BBN to the present.
Since the modulus is a bulk field, $V(r)$ should be a bulk energy
density, and $U(r) = V(r) r^n$ should have a minimum at large $r$.
Suppose that at temperature $T_*$, the modulus was
already stabilised at its minimum $r_*$. How significantly is it
perturbed as the wall fields dump energy into the extra
dimensions? We estimate this by first computing the total amount
of energy dumped into the extra dimensions. Any bulk field must
have at least a $1/M^{(n+2)/2}_{(4+n)}$ suppression for its
coupling, and so the maximum rate for dumping energy into the
extra dimensions, per unit time is
\beq
\dot{E}_{wall} -\sim \frac{T^{n+7}}{M_{(4+n)}^{n+2}} \times V_3
\eeq
where $V_3$ is the three-volume of the region of the wall losing
the energy. At worst, this energy gets entirely transfered to changing
the potential
of the radius modulus,
\beq
-\dot{E}_{wall} = \dot{E}_{rad.} = \dot{U}(r) V_3
\eeq
and so the change in $U(r)$ over a Hubble time is
\beq
\delta U(r) \sim \frac{T_*^{n+5} M_{Pl}}{M_{(4+n)}^{n+2}}
\eeq
Translating this change as $\delta U(r) \sim (\delta r)^2
U''(r_*)/2$, we obtain a bound on $U''(r_*)$ from the requirement
that $\delta r/r \lsim 10^{-1}$:
\beq
U''(r_*) r_*^2 \gsim \frac{T_*^{n+5} M_{Pl}}{M^{n+4}}
\label{softness}
\eeq
Any theory where this inequality can not be satisfied for $T_* \gsim 1$ MeV
is ruled out by cosmology during and after BBN.

Of course we do not have a theory predicting a $U(r)$ 
which naturally generates a large radius. 
Nevertheless, we can speculate on what sort of $U(r)$
can produce minima at large value $r_*$. In analogy with dimensional
transmutation, a large hierarchy can be generated if log$r$
is determined to be , say, $O(10)$. We can in any case
parametrize $U(r)$ so that
\beq
U(r) = g(r,m_{grav}) f(\mbox{log}(r m_{grav}))
\eeq
where $f(x)$ is a dimensionless function, and $g(r,m_{grav})$
has dimensions mass$^4$. One natural assumption on the form of 
$g$ is that the fully decompactified theory $r m_{grav} \to \infty$ should be a
minimum of the potential; certainly in string theory, there is a
vacuum as the string coupling goes to zero with all ten dimensions
large. In that case, it must be that $g(r,m_{grav}) \to 0$ at least as
fast
as a power law as $r \to \infty$.  We will also consider the case
where $g(r)$ is essentially flat in analogy with the ``geometric
hierarchy" potential.
The question is now whether the
theory can develop a  minimum, not for infinitely large
radius but for finite but large values of $r m_{grav}$.
Since we are interested in the
limit of a large $rm_{grav}$ anyway, we can approximate $g(r)$ at
large $r$ with its leading power law behavior
\beq
g(r,m_{grav}) \to c \frac{m_{grav}^{4-a}}{r^a}.
\eeq
Requiring $U(r)$ to be stationary then gives
\beq
\frac{m_{grav}^{4-a}}{r^{a+1}} \times (f' - a f) = 0
\eeq
so there is a minimum at a value $x_* =$log$(r_* m_{grav})$ where
\beq
\frac{f'(x_*)}{f(x_*)} = a.
\eeq
Note that the condition for the existence of a local
minimum at large $r$ is completely determined by $f$. It is
certainly not implausible that the there are dimensionless ratios
of $O(10)$ in $f$ , leading to a value of $x_*$ also of
$O(10)$, leading to a very large (but not infinite) radius.

Consider first the ``geometric hierarchy" scenario where $a=0$.
In this case,
\beq
U''(r_*) = \frac{m_{grav}^4 f''(x_*)}{r_*^2},
\eeq
and the bound from eqn.(\ref{softness}) translates to
\beq
T_* \lsim 10^{\frac{3 n}{n+5}} \mbox{GeV}
\times f''^{\frac{1}{n+5}} \times
\left(\frac{M_{(4+n)}}{\mbox{TeV}}\right)^{\frac{n+6}{n+5}}
\eeq
We do not expect $f''(x_*)$ to be larger than $O(10)$, and even if
it is larger, it is raised to small fractional power.
For $n=2$ and $M_{(6)} \sim 10$ TeV, this requires $T \lsim 100$
GeV, certainly the weakest of all cosmological bounds we have considered.
For all $n >2$, the bound is easily met with $T_* \lsim 10$ GeV
and $M_{(4+n)} \sim 1$ TeV.

The cases of intermediate ``hardness", $4>a>0$ are also less
constraining than the other cosmological bounds for obvious
reasons: $U''(r_*) r_*^2$ is enhanced by a positive power of
$m_{grav}$, so it cost significant energy to excite fluctuations
in the radius.

Finally, consider the ``soft" case of $n=4$. Here, the scale of
the potential is determined by its very size, making it ``soft"
for large radii. Here,
\beq
U''(r_*) r_*^2 = \frac{f''(x_*)-16 f(x_*)}{r_*^4}
\eeq
and $T_*$ is bounded as
\beq
T_* \lsim 10^{(\frac{3 n - 120/n}{n+5})} \mbox{GeV}
\left(\frac{M}{\mbox{TeV}}\right)^{\frac{6 + n + 8/n}{n+5}}
\eeq
Clearly for $n=2$, and even for $M_{(6)} \sim 10$ TeV, $T_* \lsim 100$
eV,and so the radius could no have settled by the time of BBN, where 
T $\sim 1$ MeV.The case $n=3$ is marginal
$T_* \lsim 1$
MeV for $M_{(7)} \sim $1 TeV, but is fine already for $M_{7} \sim 10$
TeV. For $n \geq 4$, however, even
this ``soft" scenario can be accomadated with $T_* \lsim 10$ MeV
and $M_{(4+n)} \sim 1$ TeV.

Other examples of situations where a large vev for a field can be
generated while the excitations about the minimum have a mass
uncorrelated with the vev can be constructed. In appendix 2, we
present a supersymmetric toy example of this type.

\section{The large $n$ limit}
Finally, we wish to comment on an interesting limit of our framework,
where the number of new dimensions becomes very large. This case may be
excluded
by theoretical prejudices about string theory being the true theory of
gravity,
which seems to limit $n \leq 6$ or 7, but we will ignore this prejudice here.
The $n \to \infty$ limit is interesting for many reasons. The main point
is that in this limit, the size of the new dimensions does not have to
be much larger than $M_{(4+n)}$ , solving the remaining
``hierarchy" problem in our framework. For instance, for $n=100$
new dimensions, the correct $M_{(4)}$ can be reproduced with
$M_{Pl(4+n)} \sim
2$ TeV and extra dimensional radii $\sim (1$ TeV $)^{-1}$ in size.
Since the extra dimensions are now in the TeV range, no special
mechanism is required to confine SM fields to a wall in the extra
dimensions.
Furthermore, all the KK excitations of the graviton are at $\sim 1$ TeV
and
are therefore irrelevant to low-energy physics. All the
low-energy lab and astrophysical constraints
involving emission of gravitons into the extra dimensions are gone.
Indeed,
the theory beneath a TeV is literally the SM, and all that is
required is that dangerous higher-dimension operators suppressed by a
TeV
are forbidden, a feature required even for small $n$.
The cosmology of this framework is also completely normal at
temperatures
$\lsim 1$ TeV, with no worries
about losing energy by emitting gravitons into the extra
dimensions. Apart from the usual
strong-gravitational signals at colliders, in this
limit the $KK$ excitations of SM fields for each of the $n$ new dimensions
may also be observed.

\section{Discussion and Outlook}
Over the last twenty years,
the hierarchy problem has been one of the central
motivations for constructing extensions of the SM,
with either new strong dynamics or supersymmetry stabilizing
the weak scale. By contrast, in \cite{ADD} we proposed that
the problem simply does not exist if the fundamental
short-distance cutoff of the theory, where gravity becomes
comparable in strength to the gauge interactions, is near the weak
scale. This led immediately to the requirement of new sub-mm
dimensions and SM fields localised on a brane in the
higher-dimensional space. Unlike the other solutions to the hierarchy
problem,
our scenario does not require any special dynamics to stabilised
the weak scale. On the other hand, it leads to one of the most 
exciting possibilities for new accessible physics, 
since in this scenario the
structure of the quantum gravity can be experimentally probed in the
near future. Given the amount of new physics
brought down to perhaps dangerously accessible energies, it is
crucial to check that this framework is not already experimentally
excluded.

In this paper, we have systematically studied experimental
constraints on our framework from phenomenology, astrophysics and
cosmology. Because of the power law decoupling of higher-dimension
operators, there are no significant bounds from ``compositeness" or
precision observables, which in any case do not tightly constrain
generic new weak scale physics. Rather, the most dangerous
processes involve the production of unavoidable new massless
particles in our framework--the higher dimensional gravitons--whose
couplings are only suppressed by $1/$TeV. Analogous light
4-dimensional particles with $1/$TeV suppressed couplings, such as
axions or familons, are grossly excluded. Nevertheless, we find
that for all $n>2$, the extreme infrared softness of higher
dimension gravity allows the $(4+n)$ dimensional
 Planck scale $M_{(4+n)}$ to be 
as low as $\sim 1$ TeV. The experimental limits are not trivially
satisfied, however, and for $n=2$, energy loss from SN1987A and
distortions of the diffuse photon background by the late decay of
cosmologically produced gravitons force the $6$ dimensional Planck
scale to above $\sim 30$ TeV. For precisely $n=2$, however, there
can be an $O(10)$ hierarchy between the 6-dimensional Planck scale
and the true cutoff of the low-energy theory, as was discussed in
section 3 for the particular implementation of our scenario within
type I string theory. A natural solution to the hierarchy problem
together with new physics at the accessible energies can still be
accomadated even for $n=2$.

Of course it is possible that we have overlooked some important
effects which exclude our framework. Nevertheless, these
theories have evaded the quite strong experimental limits we have
considered in a quite general way. The strongest bounds were evaded
since higher-dimensional theories are soft in the infrared. Alternately,
as $n$ grows, $r_n$ decreases and the number of available KK
excitations beneath a given energy $E$ also decreases. In fact, as
$n \to \infty$, $r_n^{-1} \to $ TeV, all KK excitations are at a TeV,
and effective low energy theory is simply the SM with no additional light
states.
This shows that this sort of new physics can not be excluded simply
because
it is exotic, the theory becomes safest in the limit of infinitely
extra dimensions.

On the theoretical front, perhaps the most important issues to
address are the mechanism for generating large radii, and early
universe cosmology. 
Other issues include proton stability, SUSY breaking in the string theory context,
and gauge coupling unification. The latter is the one piece of indirect evidence
that suggests the existence of a fundamental energy scale far above the weak scale
\cite{DG}. It has been pointed out that the existence of intermediate scale
dimensions larger than the weak scale, into which the SM fields can propagate, can
speed up gauge coupling
unification due to the power law running of gauge couplings in higher dimensional
theories \cite{Dienes}. In \cite{AADD}, the proposal of \cite{ADD} was
combined with the mechanism of \cite{Dienes} in a string context, thereby 
achieving both gravity and gauge unification near the TeV scale. 
A similar proposal was later made in \cite{BigDienes}. Alternately, \cite{Tye}
suggested that different gauge couplings may arise from different branes, leading 
to a possibly different picture for gauge coupling unification.  

Experimentally, this framework can be tested
at the LHC, and on a shorter scale, can be probed in experiments
measuring
sub-mm gravitational strength forces.
If the scale of quantum
gravity is close to a TeV as motivated by the hierarchy problem,
at least two types of signatures will be seen at the LHC
\cite{ADD,AADD}. The first involve the unsuppressed
emission of  gravitons into the higher dimensional bulk, leading
to missing energy signatures. The second involve the production of
new states of the quantum gravitational theory, such as
Regge-recurrences for every SM particle in the string
implementation of our framework. Clearly the detailed
characteristics of these signatures must be studied in
greater detail.

There is also the exciting possibility that the upcoming sub-mm
measurements of gravity will uncover aspects of our scenario.
There are at least three types of effects that may be observed.

$\bullet$ Transition from $1/r^2 \to 1/r^4$ Newtonian gravity for
$n=2$ extra dimensions. In view of our astrophysical and
cosmological considerations, which push the 6-dimensional Planck
mass to $\sim 30$ TeV, the observation of this transition will be
especially challenging.

$\bullet$ On the other hand, particles with sub-mm Compton
wavelengths can naturally arise in our scenario, for instance due
to the breaking of SUSY on our 3-brane \cite{AADD}. These will
mediate gravitational strength Yukawa forces.

$\bullet$ A new possibility pointed out in this paper is that
gauge fields living in the bulk and coupling to a linear
combination of Baryon and Lepton number can mediate repulsive
forces which are $\sim 10^6 - 10^8$ times gravity at sub-mm
distances.

\vspace{0.3cm}

{\bf Acknowledgments}: \hspace{0.2cm} We would like to thank I.
Antoniadis, M.Dine, L. Dixon, G. Farrar, L.J. Hall, Z.
Kakushadze, T. Moroi, H. Murayama, A. Pomarol,
M. Peskin, M. Porratti, L. Randall, R. Rattazzi, M. Shifman,
E. Silverstein, L. Susskind, M. Suzuki, S.Thomas and 
R. Wagoner and for useful discussions and
comments. NAH and SD would like to thank the ICTP high energy
group, and GD would like to thank the ITP at Stanford, for their
hospitality during various phases of this project. NAH is supported
by the Department of Energy under contract DE-AC03-76SF00515. SD is
supported by NSF grant PHY-9219345-004.

\section*{Appendix 1} 
In this appendix we consider the Higgs effect for the breaking of ``gauged"
translation invariance, that is spontaneous translation invariance breaking in 
the presence of gravity. In the usual KK picture (some of the) $g_{\mu
A}$ ($\mu = 0,\cdots,3$, $a =4,\cdots,3+n$
components of the higher-dimensional metric are viewed as
massless gauge fields in 4 dimensions, with the gauges\d symmetry being
translations in the extra $n$ dimensions. The result we find 
is simple and easy to as the exact analogue of
the result we found for the small mass of bulk gauge fields when the
gauge symmetry is broken on the wall. The zero modes of the 
$g^{\mu a}$ eat the $y^a$ goldstone bosons to get a mass
\beq
m^2  \sim \frac{f^4}{M_{(4)}^2}
\eeq
where $f$ is the wall tension. Intuitively, the large radius means that the
zero mode of the ``gauge field" $g^{\mu a}$ has a very small ``gauge coupling"
$\sim 1/M_{(4)}$.
The ``vev" which breaks translation invariance is nothing but the localised energy
density of the wall $f^4$, and the above formula follows. 
For completeness, however, we will consider this effect in somewhat more detail.
We will refer to the `wall' as to a
localized, stable configuration
independent of the coordinate ($x^a$),  that minimizes the action.
One can imagine the wall as some sort of topological defect in higher dimensions.

First turn off gravity and let $\Phi(x^a)$ be the vev of 
the real scalar field
forming the wall. Consider the action $S$ for the field configuration 
$\Phi(x^a + y^a(x))$. Translation invariance in $x^a$ demands that 
\beq
S[\Phi(x^a + y^a(x))] = \int d^{4+n} x f(x^a) \partial_{\mu} y^a
\partial^{\mu} y^a
+ \cdots
\eeq
where no linear term is present since $S$ is stationary at $\Phi$, and $f(x^a)$ is
some function localised around the position of the wall $x^a=0$. 
At distances large compared to the ``thickness" of the wall,
we can approximate $f(x^a) = f^4
\delta(x^a)$ where $f$ has units of mass, and 
\beq
S[\Phi(x^a + y^a(x))] \to \int d^{4 + n} x \delta(x^a) f^4 \partial^{\mu}
y^a
\partial_{\mu} y^a.
\eeq
As expected, the $y^a$ are massless dynamical degrees of freedom living on
the wall, the
Nambu-Goldstone bosons of spontaneously broken translation invariance. Global
translations in $x^a$ are realized non-linearly on the $y^a$ via $y^a(x) \to
y^a(x) + c$.  The
quantity $f^4$ can be interpreted as the tension of the wall.

Now turn on gravity, specifically the $g^{\mu a}$ ``gauge" fields
which gauge 
local translations in $x^a$:
\beq
y^a(x) \to y^a(x) + c(x), \, g^{\mu a} \to g^{\mu a} + \partial^{\mu} c.
\eeq
As usual, we can go to a unitary gauge where $y^a(x)$ are everywhere set
to
zero. In this gauge, the $g^{\mu a}$ obtain a position dependent mass term
\beq
{\cal L}_{\mbox{mass}} = \int d^{4+n} x \delta(x^a) f^4 (g^{\mu a})^2.
\eeq
That the mass term should be position dependent is intuitively obvious.
Far from the wall, no local observer knows that translation invariance has been
spontaneously broken; the graviton masses should therefore vanish
away from the wall. 

Let us expand $g^{\mu a}$ in canonically normalized KK modes
$h_{n_\alpha}^{\mu a}$,
recalling that each individual KK mode will come suppressed by $1/M_{(4)}$.  
The KK modes have already have a  mass $\sim (n/r_n)^2$, and the position
dependent
mass term from symmetry breaking becomes
\beq
{\cal L}_{\mbox{break}} = \int d^4 x \frac{f^4}{M_{(4)}^2}
(\sum_{n_\alpha} h^{\mu
a}_{n_{\alpha}})^2.
\eeq
As long as $f^2/M_{(4)}$ is smaller than $1/r_n$, the masses of the
heavy KK
excitations are not significantly perturbed by the breaking term. The zero mode 
does not have any mass in the absence of symmetry breaking, however, so it
gets a mass 
\beq
m^2_{h^{\mu a}_0} = \frac{f^4}{M_{(4)}^2}.
\eeq
Note that, for $f \sim$ TeV, this mass is $\sim ($mm$)^{-1}$, and at least for 
$n>2$ the assumption than the mass is much smaller than $r_n^{-1}$ is justified. 
For $n=2$, the first few KK modes can not be completely decoupled, and some
linear combination of them eat the $y^a$. We however still expect the
lightest graviton mode to have mass $\sim$(mm)$^{-1}$ in this case as well.

\section*{Appendix 2} 
As discussed in the text, a vague worry about having very large
dimensions comes from the impression that the potentials
responsible fro stabilizing the radius modulus will be very
``soft", and therefore the modulus will be very light,
possibly giving cosmological problems. In this example we present
an explicit counter-example to this intuition, albeit in a toy
model. We will write down a theory where
(a) field $S$ is a flat
direction to all orders in perturbation theory, b)
a potential for $S$ is generated by non-perturbative effects leading to
distinct minima very far separated
from each other, while (c) the curvature of the potential for
$S$ around its minima are completely uncorrelated
with the sizes of $\langle S \rangle$. The model is supersymmetric and
the
these features will be generated without any fine-tuning.

Consider first, an $SU(N)$ QCD with $N$
flavors $Q,\bar{Q}$ and a singlet field $S$, coupled with a
tree-level
superpotential
\beq
W_{tree} = \lambda S Tr(Q \bar{Q})
\eeq
This model has been discussed many times and has found a variety
of applications. At the classical level and to all
orders in perturbation theory, $S \neq 0$ and $Q,\bar{Q}=0$ is a
flat direction.
For $S \gg \Lambda$, $Q,\bar{Q}$ can be integrated out, and gaugino
condensation in the low energy theory gives
\beq
W_{eff}(S) = \lambda \Lambda^2 S
\eeq
This is of course the only superpotential consistent with all the
symmetries, and gives rise (at lowest order) to an exactly flat
potential
$V(S) = |\lambda \Lambda^2|^2$. Of course, the potential is modified
by corrections to the Kahler potential of $S$, and most generally
\beq
V(S) = \frac{|\lambda \Lambda^2|^2}{Z(S)}
\eeq
where $Z_S$ is the wavefunction renormalisation of $S$. For $S \gg
\Lambda$, the potential remains approximately flat since the
corrections to $Z(S)$ are perturbative, however for $S \sim
\Lambda$, this description breaks down. We are guaranteed,
however, that there is a {\it supersymmetric} minimum at $S=0$.
The exact superpotential including the quantum modified constraint
(see \cite{Review} for a review) for this case is
\beq
W = \lambda S Tr(M) + X(det M - B \bar{B} - \Lambda^{2N}).
\eeq
This superpotential admits supersymmetric vacuum with $\langle S \rangle =
0$, while the curvature of the effective potential for $S$ is $\sim
\Lambda^2$.

We can find variations on this model with copies of the
gauge group and matter to produce multiple minima for $S$.
Consider e.g. the group $SU(N)
\times
SU(N')$ with
respectively $N,N'$ flavors, and still a single singlet $S$, and
consider the tree superpotential
\beq
W = \lambda S Tr(Q \bar{Q}) + (S-m) Tr(Q' \bar{Q}')
\eeq
where $m$ is some arbitrary dimensionful scale. It is easy to see
that, for $S \ll m$, the potential looks like what we discussed
previously, with a SUSY minimum around $S=0$ with curvature $\sim
\Lambda^2$, while there is also a SUSY minimum at $S = m$ with
curvatures $\sim \Lambda'$, with a flat potential separating the
minima. In this example, the (classically flat) field $S$ can
obtain an arbitrary vev, completely uncorrelated with the
curvatures of the potential around the minima.

\def\pl#1#2#3{{\it Phys. Lett. }{\bf B#1~}(19#2)~#3}
\def\zp#1#2#3{{\it Z. Phys. }{\bf C#1~}(19#2)~#3}
\def\prl#1#2#3{{\it Phys. Rev. Lett. }{\bf #1~}(19#2)~#3}
\def\rmp#1#2#3{{\it Rev. Mod. Phys. }{\bf #1~}(19#2)~#3}
\def\prep#1#2#3{{\it Phys. Rep. }{\bf #1~}(19#2)~#3}
\def\pr#1#2#3{{\it Phys. Rev. }{\bf D#1~}(19#2)~#3}
\def\np#1#2#3{{\it Nucl. Phys. }{\bf B#1~}(19#2)~#3}
\def\mpl#1#2#3{{\it Mod. Phys. Lett. }{\bf #1~}(19#2)~#3}
\def\arnps#1#2#3{{\it Annu. Rev. Nucl. Part. Sci. }{\bf #1~}(19#2)~#3}
\def\sjnp#1#2#3{{\it Sov. J. Nucl. Phys. }{\bf #1~}(19#2)~#3}
\def\jetp#1#2#3{{\it JETP Lett. }{\bf #1~}(19#2)~#3}
\def\app#1#2#3{{\it Acta Phys. Polon. }{\bf #1~}(19#2)~#3}
\def\r nc#1#2#3{{\it Riv. Nuovo Cim. }{\bf #1~}(19#2)~#3}
\def\ap#1#2#3{{\it Ann. Phys. }{\bf #1~}(19#2)~#3}
\def\ptp#1#2#3{{\it Prog. Theor. Phys. }{\bf #1~}(19#2)~#3}

\end{document}